\documentclass[12pt]{iopart}
\usepackage{graphicx}
\usepackage{graphics}
\usepackage{epsfig}
\usepackage{subfigure}
\usepackage{amssymb}

\def\lesssim{\mathrel{\rlap{\lower4pt\hbox{\hskip1pt$\sim$}}}<}
\def\gtrsim{\mathrel{\rlap{\lower4pt\hbox{\hskip1pt$\sim$}}}>}

\newcommand{\NirHband}{$H_{150}\;$}
\newcommand{\NirKband}{$K_{200}\;$}
\newcommand{\sv}{\langle\sigma v\rangle}

\newcommand{\msun}{M_{\odot}}

\begin{document}
\title [Dark Stars: A Review]{Dark Stars: A Review }
\author[]{Katherine Freese$^{1,2,3}$, Tanja Rindler-Daller$^{3,4}$, Douglas Spolyar$^2$ and Monica Valluri$^5$\\
$^1$ Nordita (Nordic Institute for Theoretical Physics), KTH Royal Institute of Technology and Stockholm University, Roslagstullsbacken 23, SE-106~91 Stockholm, Sweden\\
$^2$The Oskar Klein Center for Cosmoparticle Physics, AlbaNova University Center, University of Stockholm, 
10691 Stockholm, Sweden\\
$^3$ Department of Physics and Michigan Center for Theoretical Physics,
University of Michigan, 450 Church St., Ann Arbor, MI 48109, USA\\ 
$^4$ Institute for Astrophysics, Universit\"atssternwarte Wien,
University of Vienna, T\"urkenschanzstr. 17, A-1180 Wien, Austria\\
$^5$ Department of Astronomy, 
University of Michigan, \\1085 South University Ave., Ann Arbor, MI 48109, USA}
\maketitle
\begin{abstract}

Dark Stars are stellar objects made (almost entirely) of hydrogen and helium, but
 powered by the heat from Dark Matter annihilation, rather than by fusion.   They are in hydrostatic and thermal equilibrium, but with an unusual power source.
 Weakly Interacting Massive Particles (WIMPs), among
 the best candidates for dark matter, can be their own antimatter and can annihilate inside the star, 
 thereby providing a heat source.  
Although dark matter constitutes only $\lesssim 0.1\%$ of the stellar mass, this amount is sufficient to
power the star for millions to billions of years.   Thus, the first phase of stellar evolution in the history of the Universe may have been dark stars.
We review how dark stars come into existence, how they grow as long as dark matter fuel persists, and their stellar structure and evolution. 
The studies were done in two different ways, first assuming polytropic interiors and more recently using the MESA stellar evolution code; 
the basic results are the same. Dark stars are giant, puffy ($\sim$ 10 AU) and cool (surface temperatures $\sim$10,000 K) objects. We follow the evolution 
of dark stars from their inception at $\sim 1 M_\odot$ as they accrete mass from their surroundings to become supermassive stars, some even reaching masses  $> 10^6 M_\odot$ and luminosities $>10^{10} L_\odot$, making them detectable with the upcoming James Webb Space Telescope.  
Once the dark matter runs out and the dark star dies, it may collapse to a black hole;
thus dark stars may provide seeds for the supermassive black holes observed throughout the Universe and at early times. 
Other sites for dark star formation may exist in the Universe today in regions of high dark matter density
such as the centers of galaxies. The current review briefly discusses dark stars existing today, but
focuses on the early generation of dark stars.

\end{abstract}



\section{Introduction}
\label{sec:Intro}

Dark Stars (DSs) are stellar objects powered by the heat from Dark Matter (DM) annihilation. 
We will focus on the DSs that may have been the first stars to form in the history of the Universe, and briefly discuss DSs
that may exist today.

The first stars formed when the Universe was roughly 200 million years old, at redshifts $z \sim 10-50$.  We will show that these first stars, which form in a dark matter rich environment,
may have been Dark Stars, powered by dark matter heating rather than by fusion for millions to billions of years.  Only after the dark matter fuel was exhausted could fusion take over as the power source inside stars\footnote{The largest supermassive dark stars 
may bypass fusion altogether and collapse directly to black holes.}.

Weakly Interacting Massive Particles (WIMPs)
 are thought to be among the best motivated dark matter candidates. Many WIMP candidates are their own antiparticles, 
 and if they are initially in thermal equilibrium in the early Universe, they annihilate with one another so that a predictable number of them remain today.  Once
 the annihilation rate drops below the Hubble expansion rate, the abundance of WIMPs freezes out.
 The relic density of these particles is approximately \cite{Jungman, BHS05}
\begin{equation}
\Omega_\chi h^2 \simeq  \frac{3 \times 10^{-27} {\rm cm}^3/{\rm sec}}{\langle \sigma v \rangle},
\end{equation}
where $\Omega_\chi$ is the fraction of the energy density in the Universe today in the form of WIMPs and
$h$ is the Hubble constant in units of 100km/s/Mpc.  With the simple assumption that
 the annihilation cross section $\langle \sigma v \rangle$ is determined
by weak interaction strength, then WIMPs automatically produce roughly the correct dark matter density today,
  $\sim 25\%$ of the total content of the Universe \cite{Ade:Planck,2013ApJS..208...19H}.  
This coincidence is known as ``the WIMP miracle" and is the reason why
WIMPs are taken so seriously as DM candidates.  The Universe as a whole consists of roughly $5\%$ baryonic material, 25\% dark matter, and 
70\% dark energy\footnote{There is some disagreement between the best fit values of the PLANCK and WMAP satellites 
\cite{Ade:Planck,2013ApJS..208...19H}, but the numbers we quote here
are roughly correct~\cite{Ade:Planck,2013ApJS..208...19H}. }.
There is a second reason for the interest in WIMPs as dark matter candidates:  WIMPs automatically exist in particle theories designed
to solve  problems that have nothing to do with dark matter.  Supersymmetric (SUSY) extensions of the standard
model of particle physics predict the existence of new partners for every particle in the standard model and, given R-parity,
the lightest of these would be dark matter candidates.  In particular, an excellent WIMP
candidate is the lightest
neutralino in the Minimal Supersymmetric Standard Model.
Models of universal extra dimensions may also have WIMP dark matter candidates in the theories (e.g. Kaluza-Klein particles)
\footnote{DM is not limited to self-conjugate Majorana states, but these are traditionally 
the most studied possibilities.}.

The same annihilation process that took place throughout the early Universe continues
  in those locations where the dark matter
density is sufficiently high for WIMPs to encounter one another and annihilate.  The first
stars to form in the Universe are a natural place to look for
significant amounts of dark matter annihilation, because they formed ``at
the right place and the right time": they formed early  (when
the Universe was still substantially denser than it is today), and at
the high density centers of dark matter halos. 

The formation of large-scale structures in the Universe -- the galaxies and galaxy clusters  -- took place via a process known as hierarchical clustering.  
As the dominant component of the mass in the Universe, dark matter drove the dynamics of this formation of structure.  Small (sub-Earth-sized) clumps formed
first; then these merged together to make larger structures; and eventually these merged yet further to produce the galaxies and clusters we see today.
These clumps of various sizes, known as ``dark matter halos," are spheroidal (prolate or triaxial) objects containing
85\% dark matter and 15\% atomic matter.  The remainder of the Universe, the dark energy, does not respond to the attractive force of gravity
and instead produces an accelerated expansion of the Universe; dark energy played no role in the formation of the first stars.

At the time of the formation of the first stars, the atomic matter in the Universe consisted only of hydrogen, helium, 
and a smattering of heavier elements (Li, B, Be) -- the products of 
primordial nucleosynthesis that took place three minutes after the Big Bang.  All
the other more complex elements were only able to form later, as the products of fusion in stars. 

Once dark matter halos about a million times as massive as the Sun ($10^6 \msun$), known as 'minihalos'\footnote{Minihalos
are defined as halos with virial temperatures below about $10^4$ K, where molecular cooling prevails over
atomic cooling mechanisms. The mass range of minihalos depends on the redshift: their maximum mass lies between 
$1-2 \cdot 10^7 M_{\odot}$ in a redshift range of z=15-20, see \cite{Shapiro04}.}, were able to form, the 
conditions were ripe for the formation of the first stars, known as Population III stars. 
 The virial temperatures of
minihalos led to molecular hydrogen cooling that allowed a protostellar cloud to start to collapse towards the center of the halo.
Reviews of the standard picture of the formation of the first stars, without taking into account the
role of dark matter, can be found in Ref. \cite{Ripamonti:2005ri, Barkana:2000fd, Bromm:2003vv, 
Yoshida:2008gn, ABN, Yoshida06}.

It was the idea of some of the authors of this review to ask, what is the
effect of the DM on these first stars?  We studied the behavior of
WIMPs in the first stars, and found that they 
can radically alter the stars' evolution \cite{DS2}. The annihilation products of the
dark matter inside the star can be trapped and
deposit enough energy to heat the proto-star and prevent it from further
collapse.  A new stellar phase results, a ``Dark Star'', powered
by DM annihilation as long as there is DM fuel, for possibly millions to billions of years.
The DM -- while only a negligible fraction of the
star's mass -- provides the key power source for the star through DM
heating. Note that the term `dark' refers to the power source, not the material or the
appearance of the star. Early DSs are stars made primarily of hydrogen
and helium with a smattering of dark matter; typically less than 0.1\% of the mass
consists of DM.  Yet, DSs shine due to DM heating. 
 
In the past few years, we have done extensive studies of the stellar structure and evolution of DSs. Dark stars are born with masses $\sim 1 \msun$ and then grow to much larger masses.
They are giant, puffy 
(10 AU), and cool (surface temperatures $\sim 10,000$K) objects
\cite{Freese08}. Since the DSs reside in a large 
reservoir of baryons
(15\% of the total halo mass), the baryons can start to accrete onto
the DSs. Our work \cite{Freese08, Spolyar09, SMDS, TRDetal14} followed the
evolution of DSs from their inception at $\sim 1 \msun$, as they accreted mass from their surroundings
 to become supermassive stars, possibly as large as $10^7 M_{\odot}$.

We now have used two different approaches in studying the evolution of dark stars.  In our initial studies, 
we assumed that the star can be described as a polytrope with the relationship between pressure $P$ and density $\rho$ at a given radius
determined by the polytropic index $n$, 
\begin{equation}\label{poly}
P=K\rho^{1+1/n},
\end{equation}
where the ``constant'' K is determined, once we know the total mass and radius \cite{chandra}. 
More recently, we have used MESA, a fully-fledged 1D stellar evolution code which allows us to solve the stellar structure 
equations self-consistently, without any a priori assumptions
on the equation of state. Remarkably, our findings
show that the previous results using polytropes roughly match (up to factors of 2) the more accurate results 
using MESA, though there are some differences in the details \cite{TRDetal14}.  Now that we have the MESA code we can
perform future studies of interesting effects like DS pulsations.

 DSs are stable as long as there is DM to fuel them.   
Indeed, as long as there is a reservoir of DM to heat the DS, the star continues to grow.  There are two main methods for sustaining 
DM fuel:  (1)  gravitational attraction of dark matter particles on a variety of orbits (extended adiabatic contraction) and (2)
capture by atomic nuclei due to elastic scattering.  The key ingredient that allows DSs to grow so much larger than ordinary 
fusion-powered Population III stars is the fact that DSs are so much cooler. Fusion-powered 
Pop III stars have much larger surface temperatures in excess of 50,000 K. They produce ionizing photons that provide a variety 
of feedback mechanisms that cut off further accretion.
In Ref.\cite{McKeeTan08}, it was estimated that the resultant Pop III stellar masses are $\sim 140 \msun$. DSs are very different from fusion-powered stars, and their cooler surface temperatures allow continued accretion of baryons all the way up to enormous stellar masses.

Supermassive DSs can in principle grow to any mass as long as the dark matter power persists --- e.g. as large as $10^7 M_\odot$ with 
 luminosities of $10^{11} L_\odot$.  Thus, they should be observable with the James Webb Space Telescope (JWST), the successor to 
 the Hubble Space Telescope,
 which is expected to be launched in 2018.  It is an exciting prospect that an entirely new type of star may be discovered in these upcoming data.

Once DSs run out of dark matter fuel, they start to collapse and can become hot enough for fusion to begin. 
Supermassive
DSs may even collapse directly to black holes.  In either case, whether or not there is  a brief period of fusion, the final end state of
most DSs will be black holes with masses in excess of $10^4 M_\odot$.  These black holes, that are the remnants of DSs, serve as seeds for the many
supermassive black holes throughout the Universe, including those at the centers of galaxies and in AGNs in the early Universe. Some may also become 
intermediate-mass black holes which have been recently discovered, see Ref.\cite{Pasham:2014, Mezcua:2015}.

For a short list of papers by various other authors that have continued the work of \cite{DS2} and explored the repercussions 
of DM heating in the first stars,
see \cite{Iocco2008, Ioccoetal08, DMfs1, DMfs2, DMfs4, Gondolo2010, Ripamonti:2010ab, SG11}. Their 
potential observability has been discussed in
\cite{SMDS, Zackrisson2010, Zackrisson2010b} (see Section \ref{sec:JWST}). The impact on reionization and the cosmic microwave background 
of populations of DSs below about 1000 $M_{\odot}$ have
been studied in \cite{Schleicher:2009} and \cite{Scott:2011}, leading to contraints on their life times and abundance 
(see Section \ref{sec:backgrounds}).

The investigation of the effect of DM on stars (in the present-day Universe) dates back to Ref.\cite{1978AJ.....83.1050S}, which studied
the impact of
massive neutrinos on galaxies and stars. 
The first paper discussing DM annihilation in stars was by \cite{krauss}.
 The first papers suggesting searches for annihilation products of WIMPs in the Sun
were by  \cite{SOS}; and in the Earth by \cite{freese86}, as well as by \cite{ksw86}. Implications of annihilation for 
stellar evolution were first explored by \cite{SS89} and \cite{bouquet}.
References ~\cite{krauss, 1985ApJ...299..994F,1985ApJ...294..663S}
first studied the effect of DM energy transport upon the Sun.  
There was a proposal to solve the solar neutrino problem  \cite{1985ApJ...294..663S, 1985ApJ...299..994F}
by using ``cosmions,'' an idea no longer pursued. 
Our Sun is 24,000 light years away from the center of our Galaxy -- too far away to be powered by DM annihilation, 
though experimenters are searching for the  products (particularly neutrinos) of the annihilation of the small amounts 
of DM captured by the Sun, Ref.\cite{Icecube2013}.

DM heating can also dramatically affect stars at the galactic center, where DM densities can be orders of magnitudes 
larger than found locally.
Ref. \cite{SS89,1989A&A...217..270B} were the first to carefully study the effects of 
dark matter on stars at the galactic center.  
More recently, the effect on today's stars has been re-examined under the assumption that DM is made of WIMPs
\cite{MW07, scott1, bertone, SFE09, CL09} or within the hypothesis of inelastic dark matter ~\cite{Hooperetal:2010}.   
Ref.\cite{Scott:jea} presented the DarkStars code: a publicly available dark stellar evolution package.
Refs. \cite{MW07,bertone} looked at DM heating in white dwarves, and neutron stars,
 at the galactic center and in globular clusters, often referred to as ``WIMP Burners."

In this review, we will focus on the effect of DM on the first stars with a brief discussion of dark stars today
in Section \ref{sec:today}.
This same WIMP annihilation process that powers DSs is also the basis for
DM indirect detection searches.   
Currently, there are several observations of excess $\gamma$-rays 
(particularly from the FERMI Gamma-Ray Space Telescope \cite{Hooper:2010mq,2012arXiv1210.5558D}) 
that may point to a dark matter annihilation origin.
Some have argued that excess positrons (seen in HEAT
\cite{Barwick:1997ig}, PAMELA \cite{Adriani:2013uda} and AMS \cite{Corti:2014ria,Accardo:2014lma}) may also 
be due to DM annihilation, but some of us have found severe bounds on this interpretation \cite{DG_AMS}; see also
\cite{Adam:2015rua}.  The origin of the positron excess is likely some other astrophysical
source.

The organization of this review is as follows: In Section \ref{sec:DMheating}, we review the three criteria required
for the initial formation of dark stars:  
 high dark matter density inside the stars, the trapping of DM annihilation products inside the stars, and 
 DM heating dominating over all
 other cooling and heating mechanisms inside the stars.  In Section \ref{sec:mass_buildup}, we discuss 
 the accretion of mass onto the
 initial DSs, thereby building up their mass.  Next we illustrate in Section \ref{sec:setup} the equilibrium structure 
 of the DSs.  We begin with the
 basic equations, and outline the polytropic approach and the MESA calculation.  We discuss the energy sources, 
 in turn dominated by DM annihilation due to adiabatically contracted DM, gravity, fusion,
 and finally captured DM.  A key ingredient in DSs is the dark matter density, so we discuss the initial density profile, as well as the adiabatic contraction 
 and capture that can bring more DM into the star.  In Section \ref{sec:stellarstructure}, we review the results of the stellar structure analysis, including 
 figures showing our results: the luminosity evolution, the Hertzsprung-Russell diagram, and the baryonic and DM 
 density profiles inside DSs.  We also
 here respond to a critique: we  point out that we agree that the initial DSs are quite dense, beyond the resolution of any existing
 simulations of protostellar collapse. Then, in Section \ref{sec:JWST} we show that JWST should be able to 
 discover supermassive DSs (more massive than $\sim 10^5 M_\odot$). As shown in Section \ref{sec:pulsations},
 a new avenue of detectability would be stellar pulsations of DSs, some having periods in the observers' frame of 
 the order of months.
 As discussed in Section \ref{sec:afterlife}, the final end product of DSs will be black holes (BHs), 
 including seeds for the supermassive BHs observed throughout the Universe and found even at early times.
  In addition, we discuss constraints on BHs with DM spikes around them residing within our Galaxy 
 today, due to $\gamma$-rays detectable by FERMI. In Section \ref{sec:backgrounds}, we review previous work on the impact of
 DSs on the cosmic microwave and infrared backgrounds. 
 A very brief review follows in Section \ref{sec:today} on dark stars in the present Universe.  
 Finally, we end with a conclusion in Section \ref{sec:summary}.

\section{Initial Formation of Dark Stars: Three Criteria for Dark Matter Heating}
\label{sec:DMheating}

Dark Stars are powered by WIMP annihilation.
The WIMP annihilation rate is $n_\chi^2 \langle \sigma v \rangle$ where
$n_\chi$ is the WIMP number density.  We take the standard annihilation cross
section (the value that produces the correct DM abundance in the Universe today) 
\begin{equation}
\label{eq:sigmav}
\langle \sigma v \rangle = 
3 \times 10^{-26} {\rm cm}^3/{\rm s}.
\end{equation}
Since the WIMP mass is converted to energy in the annihilation, 
WIMP annihilation produces energy at a rate per unit volume 
\begin{equation} \label{DMheat}
\hat Q_{DM} = n_\chi^2 \langle \sigma v \rangle m_\chi =
\langle \sigma v \rangle \rho_\chi^2/m_\chi,
\end{equation}
where $\rho_\chi$ is the WIMP mass density. As our canonical case we take $m_\chi$ = 100 GeV, but we also studied a wide range
of possible WIMP masses in the range $m_{\chi} = 10 ~\rm{GeV} -1 ~\rm{TeV}$. Since 
 the DM heating scales as
$\hat Q_{DM} \propto  \langle \sigma v \rangle/m_\chi,$
 by studying a variety of WIMP masses we are effectively studying a comparable range of annihilation
cross sections. The existence of dark stars is essentially insensitive to either the WIMP mass or the cross-section, within many orders of magnitude:
for smaller annihilation cross-sections (or equivalently larger WIMP masses), the initial DS mass would be slightly smaller, but  would
quickly grow to look similar to the DS in the case of higher cross-section or lighter WIMP mass. 

Ref.\cite{DS2} (hereafter Paper I) outlined the three key ingredients
for DSs: 1) high dark matter densities, 2) the annihilation
products become trapped inside the star, and 3) DM heating wins over other
cooling or heating mechanisms.  These same ingredients are required
throughout the evolution of the DSs, whether during its formation or during its main evolutionary phase.

{\bf First criterion: high dark matter density inside the star.}  One can see
from Eq.(\ref{DMheat}) that the DM annihilation  rate scales as WIMP density squared,
because two WIMPs must find each other to annihilate.  Thus, the annihilation is
significant wherever the density is high enough.  Dark
matter annihilation is a powerful energy source in these first stars 
because the dark matter density is high inside the early DSs.  First, DM densities in the early Universe were
higher by $(1+z)^3$ at redshift $z$. Second, the first stars form exactly in the centers of DM halos
where the densities are high. (Most of today's stars, by contrast,  are scattered throughout
the disk of the galaxy where the DM density is low; few are situated at the Galactic Center where the DM density is high).   Third, a further DM enhancement
takes place in the center of the halo: as the protostar forms, it deepens the potential
well at the center and pulls in more DM as well.  As discussed in further detail below, we have computed this
enhancement in several ways. Initially, in Paper I we used the simplistic Blumenthal method of adiabatic contraction 
(see also \cite{Natarajan:2009}), and
 found the following approximation on how the DM density follows the
(baryonic) gas density $n_h$, due to adiabatic contraction, 
\begin{equation} \label{ACanalyt}
\label{eq:simple}
\frac{\rho_\chi}{\rm{GeV} \rm{cm}^{-3}} \simeq 5 \left(\frac{n_h}{\rm{cm}^{-3}}\right)^{0.81}.
\end{equation}
We then confirmed that our results were correct to within a factor of two when we performed an exact calculation
in Ref.\cite{DS3}. We took the gas profile which was calculated in the 
first-star-simulation of \cite{ABN}; it is this profile which enters as $n_h$ in Eq.(\ref{ACanalyt}).

Inserting this result of Eq.(\ref{eq:simple}) into  Eq.(\ref{DMheat}), we find that
 WIMP annihilation produces energy at a rate per unit volume roughly given by
\begin{equation}
\label{eq:heat}
 \hat Q_{\rm DM}  \simeq  10^{-29} {{\rm erg} \over {\rm cm^3/s}} \,\,\, {\langle
    \sigma v \rangle \over (3 \times 10^{-26} {\rm cm^3/s})} \left({n_h \over {\rm
        cm^{-3}}}\right)^{1.6} \left({100~ {\rm GeV}\over m_\chi}\right) \, .
\end{equation}
In performing our subsequent calculations, we do not use this approximation but instead use adiabatic contraction to follow the DM density more accurately.
Fourth, at later stages, we also consider an additional source of DM in the star due to capture of dark matter by atomic nuclei.  
As discussed further below, as DM from the halo passes through the DS, 
some WIMPs scatter off nuclei and are captured into the DS, see Ref.\cite{FSA08, Ioccoetal08}.

We assume for
our standard case that the DM density inside the $10^6 M_\odot$ DM halo initially has an 
NFW  profile \cite{NFW}. However, our results do not depend on this choice of initial DM profile.  
Indeed, for comparison we even took the extreme case of a flat Burkert profile at the center of the halo, which is
not thought to be realistic, and still found a sufficiently enhanced DM density to produce a DS; see Ref.\cite{DS3}.
We also note that recent simulations in Ref.\cite{BZ14}
indicate that observations of cored profiles in some galaxies today could
still be consistent with cuspy early halos: supernovae and other feedback mechanisms could be responsible for reducing the 
central densities of halos today, while early halos (such as those where DSs formed) could have had NFW profiles.

{\bf Second criterion: dark matter annihilation products become trapped inside the star}.  In the early stages of first star formation,
when the gas density is low, most of the annihilation energy is
radiated away \cite{Ripamonti:2006gr}.  However, as the gas collapses
and its density increases, a substantial fraction $f_Q$ of the
annihilation energy is deposited into the gas, heating it up at a rate
$f_Q \hat Q_{\rm DM}$ per unit volume. The annihilation products and their energy spectrum depend on the WIMP model: for neutralino
DM, the annihilation products typically are $1/3$ neutrinos, $1/3$ photons and $1/3$ charged stable particles 
(electrons and positrons). While neutrinos escape from the
cloud without depositing an appreciable amount of energy, the rest can transmit energy to the core. 
Thus, we can take 
\begin{equation}
f_Q \simeq 2/3 \, .
\end{equation}
For a 100 GeV WIMP, the energy due to electrons and photons
is trapped inside the star once the hydrogen density exceeds $\sim 10^{13}$ cm$^{-3}$. Thus, around $2/3$ of the 
annihilation energy is then trapped inside the collapsing gas, 
thermalizes with it, and provides a heat source for the resulting dark star.

In our original work, we actually computed the energy deposition more precisely, as we discuss in the next few paragraphs.
The quantity $f_Q$ scales linearly with the gas density and depends on the relative
number of the various annihilation products and their energy
spectrum.  The energy spectrum of photons and electrons depends to some
extent on the exact annihilation channels.  As a typical case, we
considered spectra produced in Pythia simulations of 500 GeV
neutralino annihilation \cite{DarkSUSY,Fornengo}. Other spectral
shapes will change the precise values of our results but not the
overall effect.

Electrons below a critical energy $E_c \approx 280$
MeV in hydrogen lose energy predominantly by ionization. Higher energy
electrons do so by emission of bremsstrahlung photons. As these
bremsstrahlung photons pass near gas nuclei, they create electrons and
positrons, which in turn may generate other bremsstrahlung
photons. Thus starts a sequence of photon, electron, and positron
conversions: an electromagnetic (EM) cascade. While photons above
$\approx 100$ MeV also initiate an EM cascade, lower
energy photons transfer most of their energy to electrons in the gas
(Compton scattering).

We approximated the energy loss of electrons via ionization with 
$4.41{\rm MeV}/E $ (g/cm$^2$)$^{-1}$. For EM cascades, we 
assumed a gamma distribution for the mean longitudinal profile. 
Thus, the fraction of energy lost in traversing a thickness $X$ of gas 
equals $\gamma(a,bX/X_0)/\Gamma(a)$, where $\gamma(x,y)$ is 
the incomplete gamma function, 
$X_0=63$ g/cm$^2$ is the radiation length in hydrogen,
$a=1+b[\ln(E/E_c)-0.5]$, and $b=0.5$ \cite{rpp}.
We estimate the core thickness as $X=1.2 m_p n r_0$. Here 
$m_p$ is the proton mass, $r_0$ is the core radius, and the factor of 1.2 is 
appropriate for a uniform sphere.
We modeled the fraction of energy loss of photons by
converting each photon to an electron of the same energy after one
photon attenuation length. The latter is computed from formulas 
in \cite{rossi}, interpolated to produce the hydrogen curve in 
\cite{rpp}, figure 27.16.  There is no question that a significant fraction of the annihilation energy gets trapped inside the
star, once its hydrogen density is high enough.  Indeed, we demanded $\sim 80$ radiation lengths, which is
far more demanding than necessary to trap the energy.
We then compared the DM heating resulting from these studies with the stellar cooling.

{\bf Third Criterion: DM heating is the dominant heating/cooling
  mechanism in the star}.  We find that, for a WIMP mass of $m_\chi =
100$ GeV (1 GeV), a crucial transition takes place when the gas density
reaches $n> 10^{13} {\rm cm}^{-3}$ ($n>10^9 {\rm cm}^{-3}$).  Above
this density, DM heating dominates over all relevant cooling
mechanisms, the most important being H$_2$ cooling \cite{HM79}.

\begin{figure*} 
\begin{minipage}{0.5\linewidth}
     \centering
     \includegraphics[width=7.5cm]{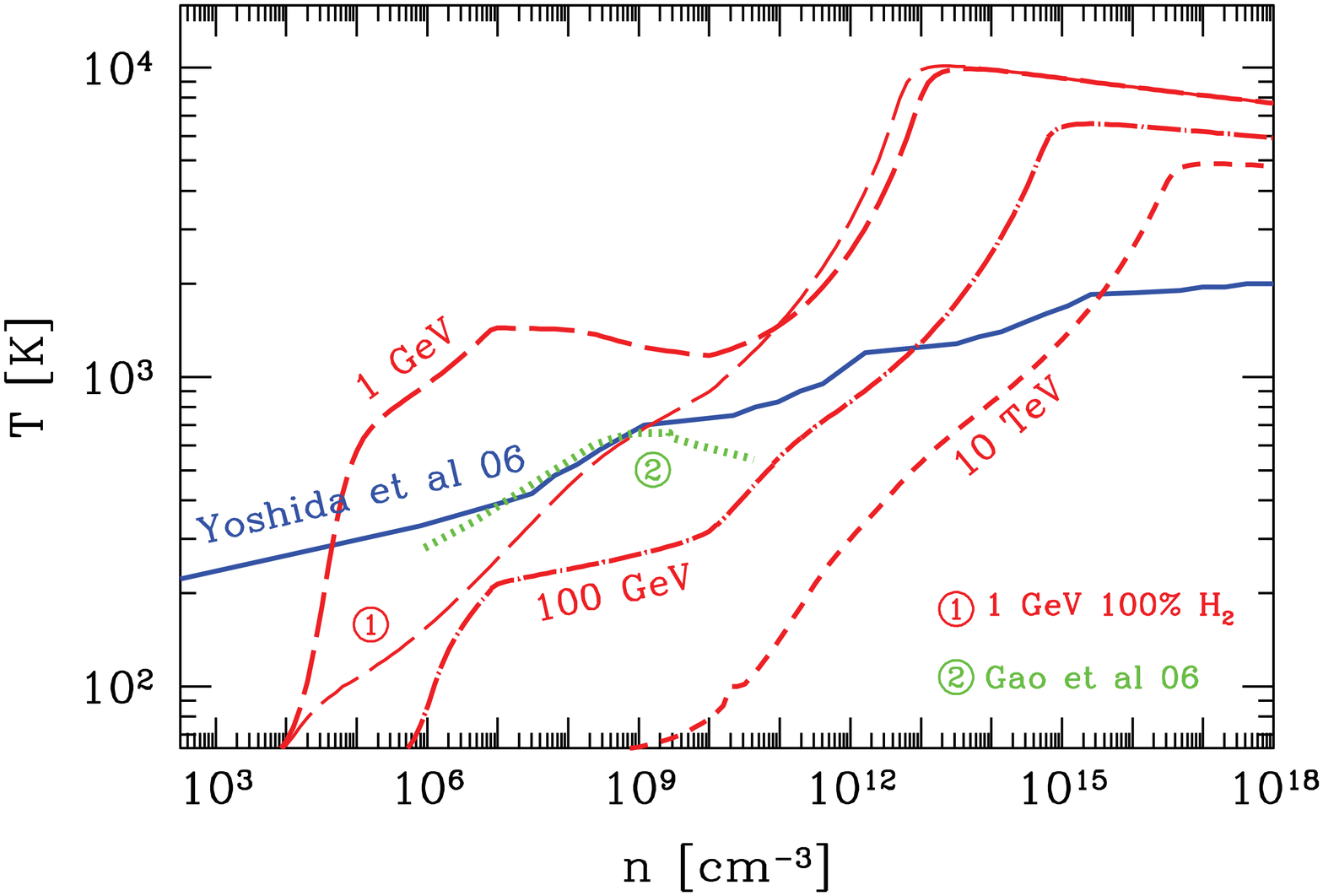}
     \vspace{0.1cm}
    \end{minipage}
    \begin{minipage}{0.5\linewidth}
      \centering\includegraphics[width=7.5cm]{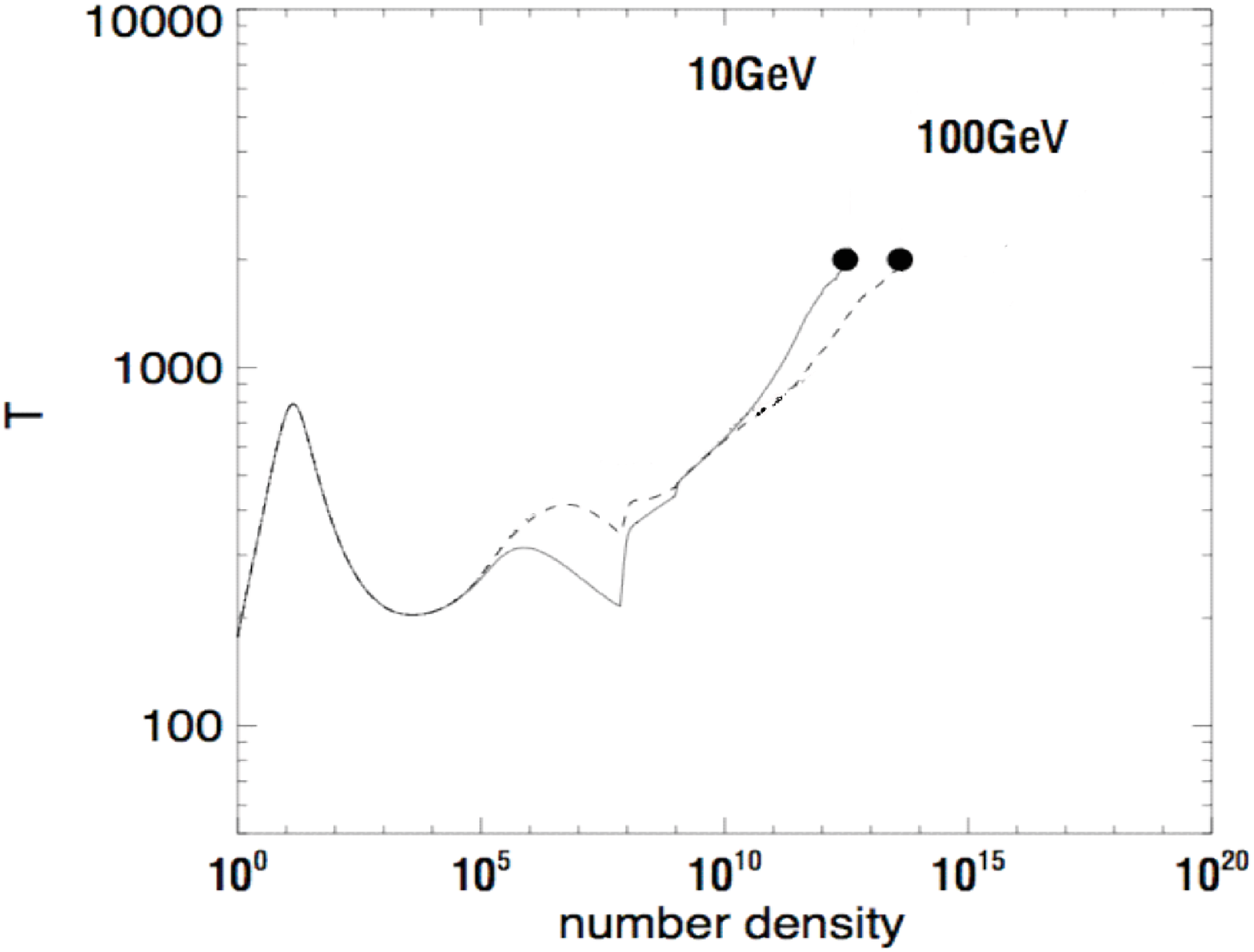}
     \hspace{0.1cm}
    \end{minipage}
\caption{Temperature (in degrees K) as a function of hydrogen density
  (in cm$^{-3}$) for the first protostars, with DM annihilation
  included. Moving to the right in the figures corresponds to moving forward in
  time. \textit{Left-hand plot:} Critical temperature lines above which DM heating dominates over cooling in the core.
  The solid (blue) and dotted (green) lines show the protostellar gas evolution from first-star simulations of 
  \cite{Gao07,Yoshida06}.
  The dashed (red) lines mark critical temperature lines for different DM particle masses, 
  (i) 1 GeV with $H_2$ density from simulations, (ii) 1 GeV assuming 100 \% $H_2$ cooling, (iii) 100 GeV, (iv) 10 TeV. At the 
  crossing point of the solid (blue) or dotted (green) with the dashed (red) lines, DM heating dominates over cooling in the
  core's evolution.
  (Figure taken from Ref.\cite{DS2}). 
  \textit{Right-hand plot:} More accurate calculation using a first-star simulation code, 
  for two different DM particle masses (10 GeV and 100 GeV). Once the ``dots'' are reached, DM annihilation dominates over H$_2$
  cooling, and a proto-DS is created. The conditions at that point are the starting values for the 
  stellar evolution calculation. (Figure courtesy of N. Yoshida, see Ref.\cite{Freese:2008be}). The results in both figures are consistent
  with each other.}
\label{fig:temp_vs_density_f1}
\end{figure*}

Figure~\ref{fig:temp_vs_density_f1} shows evolutionary tracks of the protostar in the
temperature-density phase plane with DM heating included
 \cite{Yoshidaetal08}, for two DM particle
masses (10 GeV and 100 GeV). Moving to the right on this plot is
equivalent to moving forward in time.  Once the black dots are
reached, DM heating dominates over cooling inside the star, and the
proto-DS phase begins. 
The size of the core at this point is
$\sim 17$ AU and its mass is $\sim 0.6 M_\odot$ for 100 GeV mass
WIMPs. The proto-DS keeps collapsing until equilibrium is reached\footnote{Simulations
would be required to follow the exact onset of the dark star phase in hydrostatic equilibrium. 
This is an important project for the future.}.
Eventually, above a certain baryonic density threshold, which depends on the WIMP mass, 
the annihilation products that remain trapped in the star 
 thermalize and provide a heat source for
hydrostatic and thermal equilibrium:  a new type of object is created, a DS supported by DM
annihilation rather than fusion.

One can see the power of DM heating: although the DM constitutes a tiny fraction ($<10^{-3}$)
of the mass of the DS, it can power the star (see also Section \ref{sec:stellarstructure}). The reason is that WIMP
annihilation is a very efficient power source: 2/3 of the restmass energy 
of the WIMPs is converted into useful energy for the star,
whereas less than 1\% of baryonic restmass energy is useful to a star via
fusion.

\section{Building up the Mass}
\label{sec:mass_buildup}

Once a dark star is born, with an initial mass $\sim 1 M_\odot$, it accretes mass from the surrounding medium and grows as long as the dark matter fuel persists.  
We have studied the growth of the DS by finding its equilibrium stellar structure at every step, as we build up the DS mass all the way from its initial
value to a final supermassive scale  $\sim 10^7 M_\odot$.  As the initial conditions for our simulations, we take a DS in which the baryons are  fully ionized.
We have run models for a variety of accretion rates of baryons onto the star including constant accretion rates of
$\dot M = 10^{-1}, 10^{-2}, 10^{-3}M_\odot$/yr, as well
as variable accretion rates.
In this review, we will focus for specificity on the case where dark stars are accreting matter at a 
(constant) rate of $\dot M = 10^{-3} M_{\odot}$yr$^{-1}$ in a host minihalo of $10^{6} M_{\odot}$, forming at a redshift of $z=20$,
and denoted as `SMH',
\begin{equation}
\label{eq:defineSMH}
{\rm SMH \, (Minihalo)}:  \hfil\break
 \dot M = 10^{-3} M_{\odot}/{\rm yr,} \,\,\, {\rm host \, halo \, of} \,10^{6} M_{\odot}.
\end{equation}
 
Although not discussed in this review, we also studied a second case,  labeled `LMH,' in which we considered DSs which are accreting 
at a higher rate of $\dot M = 10^{-1} M_{\odot}$yr$^{-1}$ in a larger host halo of $10^{8} M_{\odot}$ with
a formation redshift of $z=15$,
\begin{equation}
\label{eq:LMH}
{\rm LMH \, (Large \, Halo)}:  \,\,\,\,\,\,\,\,\,\,\, \dot M = 10^{-1} M_{\odot}/{\rm yr,} \,\,\, {\rm host \, halo \, of} 10^{8} M_{\odot} \, .
\end{equation}
The results for this case can be found in our series of papers for both the polytropic 
approximation \cite{Freese08, Spolyar09, SMDS}
and using the MESA stellar evolution code \cite{TRDetal14}.

\section{Equilibrium Structure of the Dark Star}
\label{sec:setup}

Dark stars are in hydrostatic and thermal equilibrium.
In our work, we adjust the DS radius 
such that the DM heating matches the radiated luminosity of the DS.  
In this section, we describe the basic equations and their implementation using both of our approaches --
our earlier polytropic models and our later 1-D stellar structure (MESA) models.

\subsection{Basic Equations}\label{basiceq}

In a numerical code for stellar structure, one key requirement is the hydrostatic equilibrium of the star. 
This is imposed at each time-step during the accretion process,
\begin{equation}\label{hydro}
\frac{dP}{dr}=-\rho(r)\frac{GM(r)}{r^2},
\end{equation}
where $P$ denotes the pressure, $\rho(r)$ is the total density and $M(r)$ is the mass enclosed in a spherical shell of radius $r$. 

{\it Polytropic approach:}
As mentioned in the introduction, in our initial studies, 
we assumed that the star can be described as a polytrope as in Eqn.(\ref{poly}),
$P=K\rho^{1+1/n}$.
We found that the energy transport is initially convective with polytropic index $n=3/2$, but as the star approaches the Zero Age Main Sequence 
(ZAMS) it becomes radiative with $n=3$. The code interpolates between $n=3/2$ and $n=3$ to account for the shift in 
energy transport as the star grows in mass. This transition happens around $(100-400) M_\odot$, after which DSs are radiative, 
following $(n=3)$-polytropes. We can determine the temperature at each point in the radial grid via the equation of state of a gas-radiation mixture,
\begin{equation}\label{eqstate}
P(r)=\frac{k_B\rho(r)T(r)}{m_u\mu}+\frac{1}{3}aT(r)^4\equiv P_g+P_{rad}.
\end{equation}
Here $k_B$ is the Boltzmann constant, $m_u$ is the atomic mass unit and $\mu=(2X+3/4Y)^{-1}$ is the mean atomic weight. 
In all resulting models $T \gtrsim 10^4$ K except near the  surface, so we use the mean atomic weight for fully ionized H and He. 
We assume a H mass fraction of X = 0.76 and a He mass fraction Y = 0.24.

At each point in the radial grid, $T(r)$ and $\rho(r)$ are used to determine the Rosseland mean opacity $\kappa$ from a 
zero metallicity table from OPAL \cite{OPAL}, supplemented at low temperatures by opacities from \cite{lenzuni} for $T<6000$ K.  
The location of the photosphere is determined by the hydrostatic condition
\begin{equation}\label{photo}
\kappa P=\frac{2}{3}g,
\end{equation}
where $g$ is the surface gravity. This corresponds to a point with inward integrated 
optical depth $\tau\sim2/3$; here the local temperature is set to $T_{\rm eff}$ and the stellar radiated luminosity is therefore
\begin{equation}\label{thermal}
L_*=4\pi R_*^2\sigma_BT_{\rm eff}^4,
\end{equation}
with $R_*$ being the photospheric radius.  The thermal equilibrium condition is 
\begin{equation}
L_* = L_{\rm tot}
\end{equation}
where $L_{\rm tot}$ is the total luminosity output from all energy sources as described below in Sec.\ref{ensources}. 

Starting with a mass $M$ and an estimate for the outer radius $R_*$, the code integrates
Eqns. (\ref{hydro}) and (\ref{eqstate}) outward from the center. The initial conditions for the central temperature and
gas density were taken from the protostellar gas profile of Ref.\cite{ABN}.
 The total luminosity output $L_{tot}$ is compared to the stellar radiated luminosity, as in Eq.(\ref{thermal}) and the radius is adjusted
 until the condition of thermal equilibrium is met (a convergence of $1$ in $10^4$ is reached).  
To re-iterate, we find that, initially, the DSs are in convective equilibrium; from $(100-400) M_\odot$ there is a transition
to the radiative ($n=3$)-polytrope case.

{\it Stellar evolution using MESA:}
In the MESA stellar code\footnote{MESA is open source software and can be downloaded from http://mesa.sourceforge.net.}, 
the equilibrium equations are built in, as are opacities and tabulated equations-of-state,
see Ref.\cite{MESAI, MESAII}. In MESA, an initial model is specified by the (stellar) mass,
a uniform composition, a luminosity, and a central temperature $T_c$ low enough to prevent nuclear burning.
Then, the total mass depends only on the central density, $\rho_c$. The initial guess for $\rho_c$ assumes an $n=3/2$ polytrope,
as for a fully convective star, although this assumption is relaxed in the subsequent search for a pre-main-sequence model,
by using MESA's routines for solving the equations of stellar structure, equation of state, and MLT (mixing-length theory) for the treatment
of convection, in order to find the $\rho_c$ for the corresponding mass. In MESA, the star grows by accreting material at a user-specified
rate, according to our choice in Eqn.(\ref{eq:defineSMH}) and (\ref{eq:LMH}). The accreted material is set to have the same entropy
than the surface layers of the model, so the accretion does not directly heat the surface, 
i.e. we adopt photospheric boundary conditions.
In order to study the effect of DM heating, we have used
 MESA's $\verb+other_energy_implicit+$ interface
to include the energy deposited in the model due to DM annihilation. This ``extra energy'' (which, in fact, is
the only energy source for our MESA dark star models) is added
self-consistently during each time step, in the same way that energy due to
nuclear reactions would be.
We note that, for our MESA models, we assume that DM is replenished due to
a continuous infall of DM on centrophilic orbits within the minihalo. Those centrophilic orbits pass arbitrarily close
through the halo center (see also Section \ref{secAC}). Thus, we adopt the same assumption as in
Ref.\cite{SMDS}, that paper on which we have based our comparison between MESA's results and polytropes.
Detailed results of our calculations and comparison to the polytropic models are published in \cite{TRDetal14}, a summary of
which is presented in Section \ref{sec:stellarstructure}.  Work is in progress to include into MESA the effects of DM capture via nuclei in the dark star, as well as nuclear fusion.
The results of this analysis will be presented in a future publication.

\subsection{Energy Sources}\label{ensources}
There are four possible contributions to the DS luminosity:
\begin{equation}
L_{tot}=L_{DM}+L_{grav}+L_{nuc}+L_{cap}
\label{eq:ltot}
\end{equation}
from annihilation of adiabatically contracted DM, gravitational contraction, nuclear fusion, and annihilation by captured DM, respectively.
The first source, heating due to DM annihilation in the course of adiabatic contraction, dominates from the time of DS formation until the DM runs out, 
\begin{equation}
L_{DM}=\frac{2}{3}\frac{1}{m_\chi} \int \rho(r)^2_{DM}\langle\sigma v\rangle dV
\end{equation}
where $m_\chi$ is the mass of the DM particle, $\rho_{DM}$ is the ambient dark matter density inside of the star 
calculated via adiabatic contraction (which will be depleted unless the loss cone is refilled, see Section \ref{secAC}.).
Here $\langle\sigma v\rangle$ is the annihilation cross section with units of cm$^3$ sec$^{-1}$.
Adiabatic contraction is discussed in \ref{secAC} and the initial DM density profile is discussed in \ref{initDM}.

Once the DM fuel runs out, the star contracts in order to maintain pressure support,
and the DS phase is over. 
The contribution $L_{grav}$ due to gravitational energy release briefly powers the star.
Then, as the star contracts, the density and temperature increase to the point where nuclear fusion begins. 
For the polytropic calculation, we included deuterium burning starting at $10^6$ K, hydrogen burning via the 
equilibrium proton-proton cycle \cite{Bahcall89}, and helium burning via the triple-alpha reaction \cite{kipp}. 

Only as the star starts to collapse (en route towards the fusion-powered main sequence), do the stellar densities become large enough to efficiently capture DM.  The captured DM then slows down the collapse and for some time can power the star. 
WIMPs from far out in the halo have orbits passing  through the star. 
The DM can then scatter off hydrogen and helium and lose enough energy to become bound to the star.  
In short order the DM thermalizes with the star (see Section \ref{sec:DMheating}). The captured DM forms a Boltzmann distribution $\rho_{cap}$ inside of the star.
 The captured DM's annihilation rate will equal the capture rate with
\begin{equation}\label{caplum}
L_{cap}=2m_{\chi}\Gamma_{cap}=2f_Q\int dV \rho_{cap}^2\sv/m_{\chi}
\end{equation}
and again $f_Q = 2/3$. The factor of $2$ reflects the fact that the energy per annihilation is
twice the WIMP mass.
Owing to the stars' more compact shape,  DSs powered by captured DM are much hotter and denser than those
powered by gravitationally captured DM.

\subsection{Dark Matter Densities}
\label{DMdens}

We turn now to the question of the dark matter density inside DSs, from their first formation in the centers
of DM minihaloes, and then forward in time as they grow in mass.

\subsubsection{Initial Profile}\label{initDM}
 The first stars form inside $\sim 10^6 \msun$ halos.  
 As mentioned above, there is still some uncertainty about the exact inner slope  of a DM 
 halo \cite{2007ApJ...667..859D,Springel:2008cc,bolshoi11}.  
  In our studies, we have used a Navarro, Frenk, \& White 
 (NFW) profile \cite{NFW}  for concreteness,
 \begin{equation} \label{iniprofile}
\rho(r)=\frac{\rho_0}{r/r_s(1+r/r_s)^2},
\end{equation}
where $\rho_0$ is the ``central" density and $r_s$ is the scale radius. 
The density scale, $\rho_0$ can be re-expressed in terms of the critical density of the Universe at a given redshift, $\rho_c(z)$ via
\begin{equation}
\rho_0=\rho_c(z)\frac{200}{3}\frac{c^3}{\rm{ln}(1+c)-c/(c+1)},
\end{equation}
where $c\equiv r_{vir}/r_s$ is the concentration parameter and $r_{vir}$ is the virial radius of the halo. 
We considered a variety of values for the concentration parameter for the polytropic case and found
little dependence.  For the MESA calculation, we chose a canonical value of $c=3.5$.

It is important to reiterate that the results of our work on DSs do {\bf not} depend on the shape of the initial density profile.
Indeed, some of us showed in \cite{DS3} that a DS results, 
 regardless of the details of the initial density profile, even for the extreme and unrealistic case of a cored 
 Burkert profile. 

\subsubsection{Adiabatic Contraction}\label{secAC}

As the baryons start to collapse
into a protostellar cloud at the center of the DM halo, the DM
responds to the changing gravitational potential well.  We
 use adiabatic contraction (AC) to describe this increase in DM
density. The baryons radiate energy and lose angular momentum. In contrast, the DM particles conserve 
certain adiabatic invariants.

We outline here the Blumenthal method (see Ref.\cite{blumenthal1,blumenthal2,blumenthal3}).
We consider the case of circular orbits (noting that it also applies to purely radial orbits)
and assume that orbits do not cross. A DM particle orbiting further out will not cross the orbit
of a DM particle further in which implies that $m(r_i)_\chi=m(r_f)_\chi$
where $m_\chi(r)$ is the mass of DM interior to a radius $r$.  Here
$r_i$ is the initial radius before the baryons fall in and $r_f$ 
is the radius of the DM particle after infall.
With the conservation of angular momentum and energy,
 $r m(r)$ is constant for each DM particle, where $r$ is the radius of the orbit. 
 The mass consists both of a dark matter component $m_\chi$ and 
 a baryonic component $m_B$ with $m=m_\chi+m_B$.
 As the baryons fall in, r must then decrease.  For instance,
 if the mass of the Sun grew slowly, the Earth's orbit would shrink.

 The Blumenthal method, despite its simplicity, is remarkably
accurate in estimating the effect of adiabatic contraction in general.
 Given an initial distribution of dark matter and baryonic matter,
 we can then find the final DM profile if we also know the final baryonic distribution,
 which we obtain from  the gas profile calculated in the 
first-star-simulation of \cite{ABN}.
By performing exact calculations for spherical halos \cite{DS3},
we found that the simple Blumenthal method \cite{blumenthal1,blumenthal2,blumenthal3} gives reliable results for the final
DM densities up to a factor of two; others have confirmed this
conclusion \cite{Natarajan:2009,Ioccoetal08, SG11}.  

In our early work, we probably underestimated the lifetime of the
DM inside the star due to AC.  At first we treated the DM
halo as spherical and ran up the DS mass to the point where the DM
initially inside the star was entirely consumed by annihilation.  The
DS mass at this point is $O(10^3)\msun$ after a lifetime of $\sim
300,000$ years, and the amount of DM consumed has only amounted to
$\sim 1 \msun$.  In a spherical DM halo, the orbits of DM particles
are planar rosettes \cite{BT08}, conserving energy as
well as all three components of angular momentum; consequently the
central hole (or ``empty loss cone'') that results from DM
annihilation cannot be repopulated once it is depleted. However,
it is well known that DM halos formed in hierarchical structure
formation simulations are not spherical but are prolate-triaxial
\cite{Bardeenetal86, Dubinski:1991, Jing:2002,Bailin:2005,Allgood:2006} with typical axis ratios of 
(short-axis)/(long-axis) $\sim 0.6-0.8$. In triaxial potentials, the
orbits do not conserve any component of angular momentum. In particular, there are two
families of ``centrophilic orbits'' (box orbits and chaotic orbits)
which oscillate back and forth through the potential and can travel
arbitrarily close to the center \cite{Schwarzschild79, GS81, Gerhard:1985,de-Zeeuw:1985,Schwarzschild93, 
Merritt:1996, Merritt:1996a}.
These orbits have the low angular momentum necessary for
 replenishing DM in the center, implying that the duration of a DS phase 
 could increase by many orders of magnitude, from $\sim 300,000$ years (see Section \ref{sec:stellarstructure} and Figure \ref{fig:f2})
 to a timescale many orders of magnitude longer,
 depending on the cosmological evolution of the minihalo hosting the DS. 
 
 Several studies of the shapes of DM halos have shown that when baryons condense into the center of the DM halo, the shape of the 
 inner region (0.3 $\times r_{\rm vir}$) becomes nearly spherical 
 \cite{Dubinski94, Ketal04, Detal08}. This raises the concern that the formation of the DS itself would 
 lead the DM halo to convert from being triaxial back to being spherical before the
  DS reaches supermassive size. In a series of papers,  Ref.\cite{Valluri10, Valluri12} carried out a detailed statistical analysis of the orbital properties of DM particles in triaxial and prolate-triaxial halos 
  in which baryonic components were grown adiabatically. They analyzed over 20,000 randomly selected halo orbits as a function of radius using an automatic orbit classification tool that relies on the 
  fundamental oscillation frequencies of each orbit (using frequency mapping), see Ref.\cite{Laskar90, VM98}.  
  It has been shown in \cite{Valluri10, Valluri12} that the orbits of DM particles remain 
  primarily centrophilic (``box'' orbits and chaotic orbits) in the inner-most regions of the DM halo despite the fact that the 
  halos become quite close to (but not exactly) spherical. More precisely, it has been shown in \cite{Valluri10} that over 50 \%
  of the orbits overally are centrophilic which, after a few orbital periods, will get arbitrarily close to the center of
  the potential. More recently, 
  \cite{Valluri13} analyzed DM halo particles in a fully cosmological hydrodynamical simulation of a disk galaxy, 
  part of the McMaster Unbiased Galaxy Simulations \cite{Stinson10}. They 
  found that, in the fully cosmological simulations, DM halos became even more spherical than when the 
  baryonic components were grown adiabatically, nonetheless an even more significant fraction of DM halo 
  particles remained on box and chaotic orbits, indicating that, even when the full hierarchical merger history of a halo is 
  considered, DM particles will continue to be on orbits that penetrate to the central cusp, 
  thereby continuing to provide heat for the DS.
 
 As a consequence of the prolonged period of dark matter heating, supermassive DSs (SMDSs) can result, with masses up 
 to $10^5-10^7 \msun$ and luminosities up to $10^8-10^{11} L_\odot$. 
 The amount of dark matter required inside the star to sustain long enough DM heating to reach these stellar masses is still small;
 e.g.  $\sim 100\msun$ for a $10^5 \msun$ SMDS for an
accretion rate of $\dot M = 10^{-1}\msun$/yr and $\sim 10^4 \msun$ for an accretion rate of $\dot M = 10^{-3}\msun$/yr, 
out of a total $10^6\msun$ halo.

 In reality, DSs will form in a variety of  environments and experience mergers during their life times. Such mergers could disrupt or 
 enhance the growth of SMDSs at various stages, resulting in DSs of a variety of masses.

\subsubsection{Dark Matter Capture}
\label{dmcapture}

In the previous subsection, we have only discussed the dark matter brought into the DS purely by gravity
via adiabatic contraction.
However, the DM reservoir inside the DS can be refueled by DM capture, in addition. This refueling requires
an additional piece of particle physics: scattering of DM off the nuclei inside the star.  This is the same elastic scattering that 
is the origin of searches for detectable signals in dark matter ``direct detection experiments," such as DAMA, Xenon-based detectors,
 SUPERCDMS, CRESST, PICO, and many others.
 The role of captured DM in DSs was first
noticed simultaneously in Ref.\cite{Iocco2008, FSA08}.
The capture rate is sensitive to two uncertain quantities: the scattering cross section of WIMP interactions with the nuclei and  the background DM density.  

The cross section for elastic scattering depends on the coupling of DM to ordinary matter.  The two cases most often
considered are ``spin-independent"  (SI) and ``spin-dependent" (SD) scattering.  For WIMP masses above $\sim$ 1 GeV,
the constraints on spin-independent scattering are much stronger than for spin-dependent scattering. Hence, we consider here
the optimal case for capture in DSs, namely SD scattering.
   The PICO experiment \cite{PicoTalk} has preliminary constraints which are the most restrictive for SD scattering,
 $\sigma<10^{-39}$ cm$^2$ for $m_\chi=30$ GeV (and much weaker for other masses). For $m_\chi\sim 1$ GeV, the
 bound is roughly $\sigma<10^{-37}$ cm$^2$, see Ref.\cite{Archambault:2012pm,Aprile:2013doa,PicoTalk}.
 \footnote{Spin-independent scattering has much stronger constraints.
 For $m_\chi\sim1$ GeV, CDMSlite~\cite{Agnese:2013jaa} constrains the cross section to be less than roughly $10^{-40}$ cm$^2$.
 For $m_\chi=30$ GeV, LUX  constrains the cross section to be less than $4\times10^{-46} $ cm$^2$ \cite{Akerib:2013tjd}.
 Colliders (assuming effective operators with a large cut off) can also in some cases provide constraints; 
 though the validity of the effective operator approach is not general ~\cite{Haisch:2012kf,Fox:2011pm}.  In those cases where the effective operator approach is valid,
CMS and ATLAS constrain the spin-dependent cross sections to $\sigma<10^{-40}$ cm$^2$ for $m_\chi<100$ GeV and this 
 weakens to $\sigma<10^{-39}$ cm$^2$ for $m_\chi$ above a few hundred GeV.  
Also, for energies (DM masses, respectively) above the cut off mass - given by the ratio of mediator mass to coupling constant -, 
the constraints should not be trusted, since an effective operator approach is invalid.}

In the interesting case of a high DM density environment, the additional
DM power due to captured DM can allow the DS to keep growing, to the point where SMDSs of mass $> 10^5 \msun$ can 
result, see Ref.\cite{SMDS}.   
Thus, SMDSs can arise due to two separate DM populations: the DM brought in via gravity (extended AC) or the DM brought
in via capture. The two different types of SMDSs are quite different objects, as we show below.

\section{Results of Stellar Structure Analysis}
\label{sec:stellarstructure}

In this section, we discuss the results of our stellar structure analysis for
a variety of cases.  We show results for both the case of ``extended AC " (where 
DM is brought into the star via gravity only), as well as the case ``with capture" (where DM is captured in the star via elastic scattering).  We consider a range
of  WIMP masses $m_\chi$ = 10 GeV, 100 GeV, and 1
TeV.  We show results using our two approaches:  polytropic stellar models and models obtained with the MESA stellar evolution code. 
In this section, we use the accretion rate and halo size defined in Eqn.(\ref{eq:defineSMH}); for other cases our results can be
found in our series of papers.
Figures 2-5 show the stellar models that are the result of our analysis.  Figures 2 and 3 were obtained 
using polytropic stellar interiors \cite{SMDS}, while Figures 4 and 5 are our results using the MESA stellar
evolution code \cite{TRDetal14}.

\begin{figure}
\centering
\includegraphics[scale=0.45]{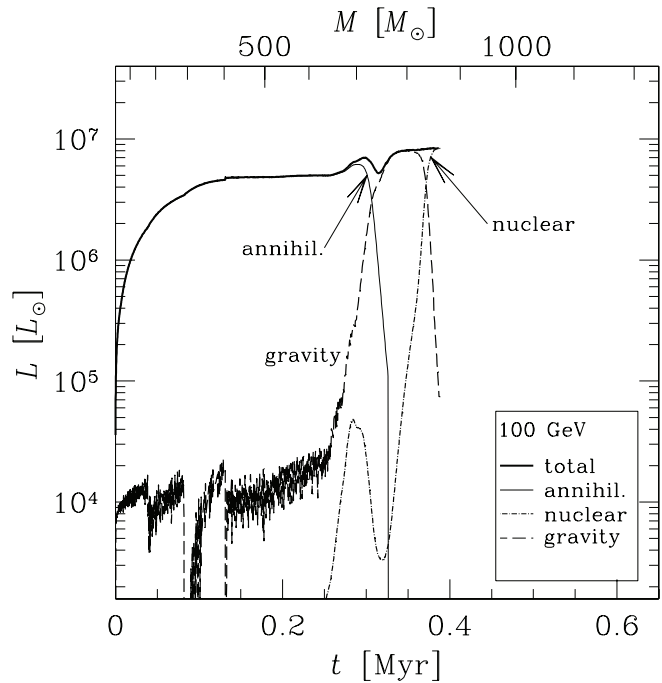} 
\caption{Luminosity evolution of dark stars for the case of a 100 GeV WIMP mass as a function of time
({\it lower scale}) and stellar mass ({\it upper scale}). Results were obtained assuming polytropic interiors for the DS.  The solid 
top curve is the total luminosity.  The lower curves give the partial contributions of  different sources of energy 
powering the star  without capture. The total luminosity is initially dominated by DM annihilation 
(the total and annihilation curves are indistinguishable until about 0.3 Myr after
 the beginning of the simulation); then  gravity dominates, followed by nuclear fusion. In this plot, we have cut off the DM heating by
 hand at 0.3 Myr when the DS mass approaches 850 $M_\odot$, as one example of a final possible DS.  In reality, the DM annihilation continues as long
 as the DM reservoir is replenished (depending on the DM environment of the particular star); thus the DS could continue to grow
 for a longer time and the final DS mass could be many orders of 
 magnitude larger than in the plot shown here.  The purpose of this figure is to show the full evolution of an illustrative 
 example of a dark star and its many contributing heat sources. (Figure taken from Ref.\cite{Spolyar09}).}
\label{fig:f2}
\end{figure}

Figure~\ref{fig:f2} shows the luminosity evolution of a DS for the case of adiabatically contracted DM only (no capture) for polytropic interiors
and assuming that
the DM runs out when the star reaches about 800 $\msun$ at roughly 0.3 Myr after the beginning of the simulation.  
This cutoff in the lifetime of the DS phase was obtained using the overly conservative assumption
that DM would run out inside the DS at that time (based on the incorrect assumption 
that the DM halo is spherical; see the discussion in Section \ref{secAC}). 
 In reality, the DS could keep growing for a much longer time,
depending on its DM environment. The various contributions to the luminosity in Eq.(\ref{eq:ltot}) are plotted
as well as the total luminosity (solid curve).  The total luminosity is initially dominated by DM annihilation; then  gravitational contraction dominates, followed by nuclear fusion.

\begin{figure}
\centering
\includegraphics[scale=0.35]{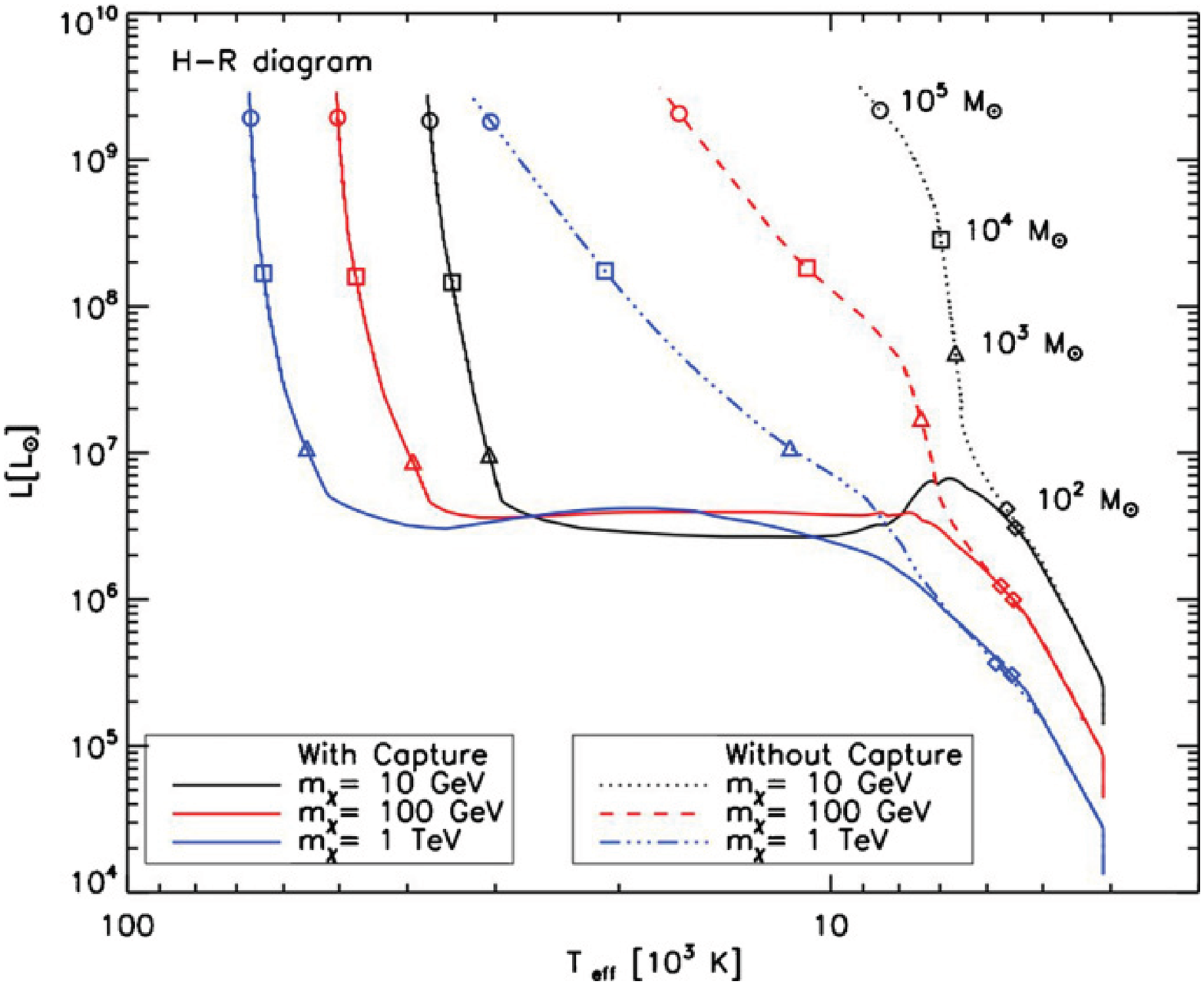}
\caption{Hertzsprung-Russell diagram for DSs forming in SMH, as defined in eq.(\ref{eq:defineSMH}), and a variety of WIMP masses as
  labeled for the two cases: (i) ``without capture'' but with extended
  adiabatic contraction (dotted lines), assuming no significant depletion of DM due to annihilation,
  which is equivalent to assuming a replenishment of DM due to centrophilic orbits, as explained in text; (ii) ``with capture''
  (solid lines).  Results were obtained assuming polytropic interiors for the DS.
  The case with capture is for product of scattering
  cross section times ambient WIMP density $\sigma_c \bar\rho_\chi =
  10^{-39} {\rm cm}^2 \times 10^{13}$GeV/cm$^3$ (the maximum allowed cross section for all WIMP masses and the 
  maximum reasonable ambient density for 100 GeV WIMPs).
  Given these values, DSs must become dense enough for DM capture to happen.
   This explains the horizontal lines
  in the evolution of the case ``with capture''.
  Labeled are also
  stellar masses reached by the DS on its way to becoming
  supermassive. The final DS mass was taken to be $1.5\times 10^5
  \msun$ (the baryonic mass inside the initial halo), but could vary
  from halo to halo, depending on the specifics of the halo mergers
  (Figure taken from Ref.\cite{SMDS}).
}
\label{fig:dens_f3}
\end{figure}

The Hertzsprung-Russell diagram resulting from our studies using polytropic interiors in Ref.\cite{SMDS} is shown in
Figure \ref{fig:dens_f3}.  The DS travels up to increasingly
higher luminosities as it becomes more massive due to accretion.   As
the mass increases, so does the surface
temperature. In the cases
``with capture'', we have taken the (overly conservative) assumption
that the DM from adiabatic contraction is depleted after $\sim
300,000$ yrs as in our earlier papers; then the luminosity plateaus
for some time, while the DS contracts until, eventually, it is dense enough
to capture further DM.
  
\begin{figure}
\centering
\includegraphics[scale=0.45]{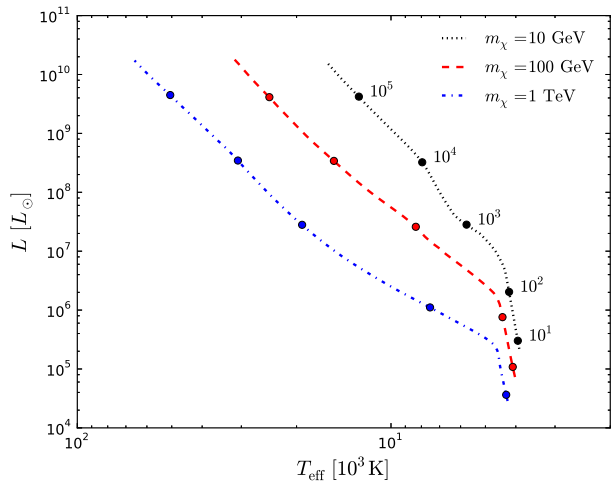}\\
\hspace{0.5cm}
\caption{Hertzsprung-Russell diagram for DSs forming in SMH (defined in eq.(\ref{eq:defineSMH})) and a variety of WIMP masses,
using the MESA stellar evolution code.
The calculations assume extended adiabatic contraction and no significant depletion of DM due to annihilation.
DM capture is not considered. Labeled are
  stellar masses reached by the DS on its way to becoming
  supermassive; see also Ref.\cite{TRDetal14}.
}
\label{fig:HRMESA}
\end{figure}

We note that, for the case ``without capture'', the tracks in the H-R diagram are
unchanged by varying the accretion rate: only the time it takes to get
from one mass stage to the next changes, but the curves we have
plotted apply equally to all accretion rates. 
The Hertzsprung-Russell diagram obtained with MESA can be found in Fig.\ref{fig:HRMESA} for the case
of extended AC.  DM capture and fusion 
have not yet been included, and work is in progress to implement these as well. 
In general, we found remarkably good overall agreement with the basic results of the polytropic models; however there are some
differences. Using MESA, we found that, in the mass range of $10^4-10^5 M_{\odot}$, our DSs are hotter by a factor of $1.5$ than those
in \cite{SMDS}, are smaller in radius by a factor of $0.6$, denser by a factor of $3-4$, and more luminous by a factor of $2$.
Thus, the overall colors of our DSs are not very different from the polytropic models, while the higher luminosities we find improve
the prospect of observability of DSs with upcoming space telescopes, such as the JWST.

In the HR diagrams of Figure \ref{fig:dens_f3} and \ref{fig:HRMESA}, the curves with higher values of WIMP mass $m_\chi$ lie to the left of
the curves with lower $m_\chi$.  This can be understood as follows.
The DM heating rate in Equ.(1) scales as $Q \propto \rho_\chi^2 /
m_\chi$.  Hence, to reach the same amount of heating and achieve the
same luminosity at higher $m_\chi$, the DS must be at higher WIMP
density, i.e., the stellar radius must be smaller, the DS is hotter,
and the corresponding surface temperature $T_{\rm eff}$ is higher.  Also,
for higher $m_\chi$ the amount of DM in the star is smaller since the
star is more compact for the same number of baryons, but $\rho_\chi
\propto n^{0.81}$, see Equ.(\ref{eq:simple}).

\begin{figure*} 
\begin{minipage}{0.5\linewidth}
     \centering
     \includegraphics[width=7.5cm]{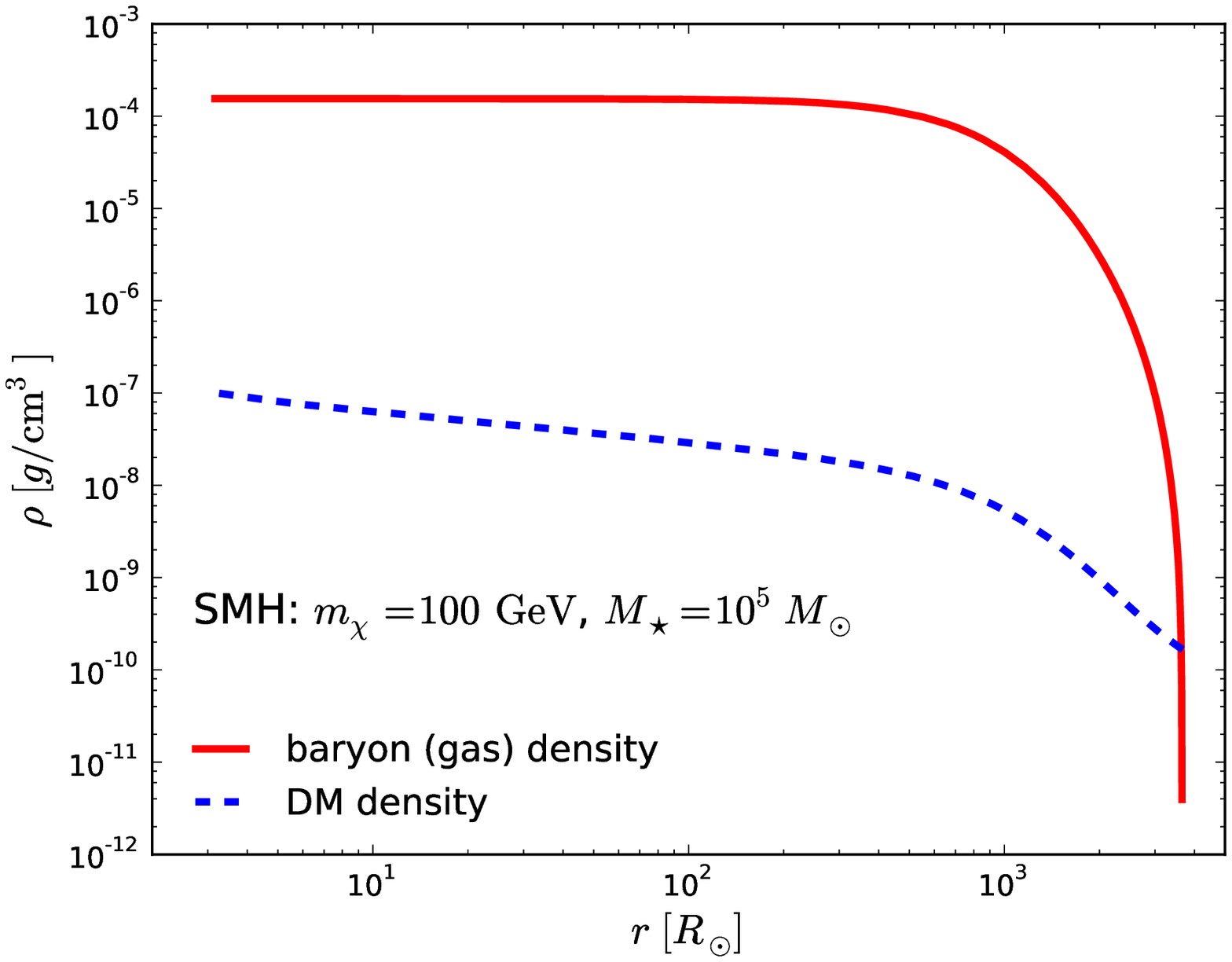}
     \vspace{0.1cm}
    \end{minipage}
    \begin{minipage}{0.5\linewidth}
      \centering\includegraphics[width=7.5cm]{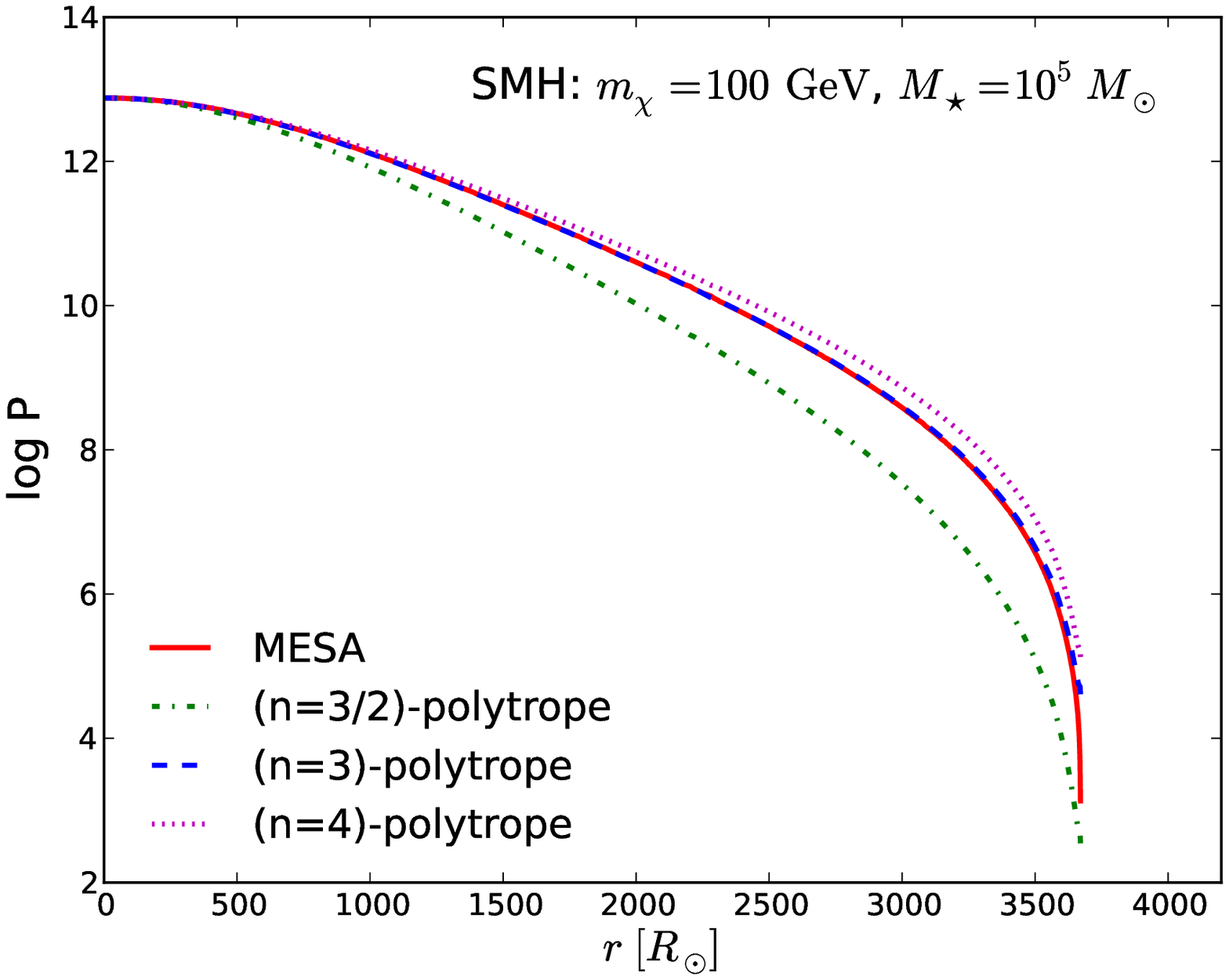}
     \hspace{0.1cm}
    \end{minipage}
\caption{A dark star with mass $10^5 M_{\odot}$ forming in SMH and WIMP mass 100 GeV: results were obtained using the MESA stellar
 evolution code. {\it Left-hand plot:}
 Baryonic gas and DM density profiles within the DS. The DM density is roughly three orders of magnitude below the gas density. {\it Right-hand plot:} 
Stellar total pressure as a function of stellar radius: 
comparison of MESA's results with polytropes
of index $n=3/2$, $n=3$ and $n=4$, assuming the same central pressure and density. The figure shows that supermassive DSs can be
very well approximated by $(n=3)$-polytropes, confirming earlier results. Note the different $x$-axis scale in the two figures;
see also Ref.\cite{TRDetal14}.}
 \label{fig:DSMESA}
\end{figure*}

The density and pressure distribution within a supermassive DS of $10^5 M_{\odot}$ obtained using the MESA code
can be found in Figure \ref{fig:DSMESA}.
As expected from earlier results, the DM density is roughly three orders of magnitude below the baryonic gas density, i.e.
the DM mass contribution in DSs is very small, see left-hand-plot. For the example shown in  Fig.\ref{fig:DSMESA}, the total mass in
DM only amounts to about $20 M_{\odot}$, or $0.02\%$ of the DS mass. The shape of both density profiles as well as their absolute magnitude agrees
excellently with the results in Fig.3, case 1 of Ref.\cite{Spolyar09}.
 The right-hand-plot of Figure \ref{fig:DSMESA} shows the total pressure inside the DS as a function of
radius. To this end, we plot MESA's accurate result together with the known run of polytropes of index $n=3/2$, $n=3$ and 
$n=4$, respectively.
Except close to the surface, we can see that supermassive DSs can be very well approximated by $(n=3)$-polytropes, as 
has been found in the earlier work of Ref.\cite{Spolyar09, SMDS},
and as we have described above. In fact, our MESA results confirm even the overall evolution of the internal pressure distribution and 
energy transfer of DSs on their way of becoming supermassive, as follows. While MESA does not need to rely on the assumption of
polytropic equations-of-state, we can still define an ``effective polytropic index'' via
\begin{equation} \label{neff}
n_{\rm{eff}} = \left[\frac{\log (P/P_c)}{\log (\rho/\rho_c)} - 1 \right]^{-1},
\end{equation}
where $P$ and $\rho$ are the exact MESA values for the pressure and density, with $P_c$ and $\rho_c$ their
values at the DS center.  For low-mass DSs with a stellar mass around $(10-20) M_{\odot}$, $n_{\rm eff}$ is close to 
the value of $3/2$, appropriate for a fully convective star. This value steadily increases to above $n_{\rm eff} = 2$
for more than $100 M_{\odot}$. Around this point, the luminosity due to radiation transfer starts to be of the order of the
luminosity due to convection, i.e. $L_{rad} \sim L_c$. As the DS continues its mass growth, $n_{\rm eff}$ continues to approach
a value of $3$, and $L_{rad}$ becomes increasingly important. The energy transport in supermassive DSs is thus dominated by
radiation transfer.

In a beautiful paper in 1963, Hoyle and Fowler \cite{HF86} studied supermassive
stars in excess of $10^3 \msun$ and found results germane to our
work. They treated these as $(n=3)$-polytropes, dominated
by radiation pressure, and found the following results: $R_* \sim
10^{11} ({M_* / \msun})^{1/2} (T_c/10^8~{\rm K})^{-1} {\rm cm}$, ${L_*
  / L_\odot} \sim 10^4 ~{M_* / M_\odot}$, and $T_{\rm eff} \sim 10^5
(T_c/10^8{\rm K})^{1/2}~{\rm K}$.  While some of the details of their
calculations differ from ours, taking the central temperature
appropriate to DSs in the above relations roughly reproduces our
results (to O(1)). By using the temperature appropriate
to DSs with extended AC ($T_c \sim 10^6$K) rather than the much
higher central temperature ($T_c > 10^8$K) appropriate to nuclear
power generation, the above relations show that DSs have much
larger radii and smaller surface temperatures than fusion-powered
stars. The fact that DSs are so fluffy objects makes them resilient against
general-relativistic instabilities, since general-relativistic corrections, which
would act in the direction of destabilizing supermassive stars, scale as
$G M_*/R_*$. The upper limit on the allowed stellar mass is thus larger for DSs.
    
For the cases with capture, we take $\sigma_c \bar\rho_\chi =
  10^{-39} {\rm cm}^2 \times 10^{13}$GeV/cm$^3$ (the product of the maximum allowed scattering 
  cross section from direct detection experiments, and the maximum reasonable ambient density for 
  100 GeV WIMPs). With these parameters,
  DSs must become dense enough for DM capture to happen.     
The cases with capture
all take place at higher stellar densities than the cases without; the
density must be high enough to be able to capture WIMPs.  Consequently,
the surface temperature is larger and accretion is shut off more
easily by radiation coming from the star. Between 50,000 K and 100,000 K feedback effects were
included in the polytropic case, and they act to reduce the accretion rate, but they never
shut it off entirely for densities above $5 \times 10^{10}$ GeV/cm$^3$.
In reality, a star that is moving around can sometimes hit pockets of
high $\bar\rho_\chi$ (where it is DM powered and grows in mass) and
sometimes hit pockets of low $\bar\rho_\chi$ (where fusion takes
over).  As long as the ambient density remains at least this large,
the star can reach arbitrarily large masses and eat the entire
baryonic content of the halo.
  
The capture mechanism depends on
 the product of scattering cross section times ambient
density, $\sigma_c \bar \rho_\chi$, rather than on either of these
quantities separately.  Hence, our current discussion
could trade off ambient density vs. cross section. For example, the
product is the same for $\bar \rho_\chi = 5 \times 10^{10}$ GeV/cm$^3$
and $\sigma_c = 10^{-39}$ cm$^2$ as it is for $\bar \rho_{\chi} =
10^{13}$ GeV/cm$^3$ and $\sigma_c = 5 \times 10^{-42}$ cm$^2$.  
  
Above $\sim 100 \msun$, one can see that the stellar luminosity scales
as $L_* \propto M_*$ and is the same for all models for a given
stellar mass; this statement is essentially true for all stars no
matter the power source.  The reason is that at these masses, the star
is essentially radiation pressure supported throughout. This same scaling in supermassive stars
was already noticed
in Ref.\cite{HF86}. We confirmed with MESA that the luminosity of DSs stays comfortably below the Eddington limit luminosity (which scales as $L_{\rm Edd}
\propto M_*$) for most of the DS evolution
and tracks it just below at the supermassive end. This and other details of the stellar structure
of the first DSs will be presented in \cite{RDF}.
  
   The caveat in the above discussion is that in order to continue to grow, the DM orbits must continue to penetrate 
   into the middle of the DS for a greater length of
time; it is the DM heat source that keeps the DS cool enough to allow it to
continue to accrete baryons.  Additionally, the assumption that
baryons continue to accrete onto the DS must continue to hold.  Yet,
in the time frame required, the original $10^6\msun$ minihalo will merge
with its neighbors, so that both the baryon and DM densities are
disturbed. These mergers could affect the DS in one of two ways:
either they provide more baryons and DM to feed the SMDS so that it
ends up being even larger, or they disrupt the pleasant high DM
environment of the SMDS so that it loses its fuel and converts to an
entirely fusion-powered star.  Continued growth of the DS is quite
plausible since simulations with massive BHs in mergers show that they
prefer to sit close to the center of the density distribution, or find
the center in a short time after the merger. 

Someday detailed cosmological simulations will be required to answer
this question. Individual DSs in different halos may end up with a
variety of different masses, depending on the details of the evolution
of the halos they live in.  The case studied in \cite{SMDS} and \cite{TRDetal14} is clearly
a simplistic version of the more complicated reality, but illustrates
the basic idea that supermassive stars may be created by accretion
onto DSs, either with or without capture.  
  
\subsection{Response to criticism}

Before we move on to describe the observable properties of dark stars, 
we would like to address some recent critiques of the
feasability of a DS phase.  The concern is whether DSs are ever able to form at all; we feel that the issues raised in the 
criticism are easy to address. We refer to a note on the arXiv for a more quantitative response, see \cite{Gondolo13}.

First, Ref. \cite{Ripa10} and \cite{Smith12} have performed simulations of collapsing protostellar clouds and 
 noted that the collapse continues past the hydrogen density quoted in \cite{DS2} as the point where DM heating starts to dominate. 
 This fact led to the incorrect conclusion 
  that DM annihilation is not potent enough for the establishment
  of a DS in hydrostatic equilibrium, powered by DM heating. However, 
  there is in reality no disagreement between the results of these simulations and 
  the existence of a dark star.  Once the DM power dominates, the
  protostellar object must indeed collapse further before it becomes a real star, in hydrostatic
  and thermal equilibrium.  Specifically, the objections have arisen because for 100 GeV WIMPs, 
  the simulations of \cite{Ripa10} and \cite{Smith12} found that the collapse continues past a hydrogen 
  density $n_h=5 \times 10^{14}{\rm cm}^{-3}$, which is the limit of their simulations.
Indeed, we agree in \cite{Freese08} and
  \cite{Spolyar09} that the cloud continues to collapse past this point; and we find that the initial
  dark star forms later, when the hydrogen density is  $n_h \sim 10^{17} {\rm cm}^{-3}$, three orders of magnitude 
  higher than the limits of the simulations. The aforementioned simulations are unable to reach densities
  this high, and are therefore unable to directly address the dark
  star regime.  A very interesting project would be to push the simulations further to observe the
actual formation of the dark star.

    Second, concern has been addressed in Ref.\cite{Stacy14} 
that the DM accessible to the DS at the centers of minihalos may run out rapidly enough, so as to limit the DS phase
to only a few thousand years. However, as we have described in Section \ref{secAC}, typically half of the dark matter orbits come in from far outside the 
small region near the halo center that is studied in these simulations; these dark matter particles do indeed continue to 
replenish the DM required for DSs to grow.  

Third, a growing body of literature finds that the accretion disk around the first protostars can fragment, leading to the 
possible formation of multiple systems and removal of the central object from the DM cusp, see e.g.
Ref.\cite{Stacy10, Clarketal2011, Greifetal2012}.
However, the inclusion of DM heating in Ref.\cite{Smith12} resulted in a stabilizing effect, preventing fragmentation around the
central protostar. Future studies with high enough resolution to capture the mutual dynamical effects of baryons and DM in these
central regions will be needed.

Finally, in existing simulations in the papers criticizing our work, there are inaccuracies in the energy injection
that should be corrected:
the prescriptions for the treatment of the effect of DM annihilation energy injection into the 
primordial protostellar gas are valid for particle energies in the keV range, which are lower than those expected from WIMP decay. 
We note that in our original work in \cite{DS2}, we required
that the WIMP annihilation products have at least 80 radiation lengths in the DS - this is overkill in ensuring that they
will be trapped and heat the protostar. Our work did take into account the electromagnetic cascades 
experienced by WIMP annihilation products with energies exceeding 100s of MeVs. It is the electrons with energies above around 
280 MeV and photons above around 100 MeV which initiate a further 
electromagnetic cascade in their hydrogen environment. Lower-energetic electrons lose their energy pre-dominantly by ionization,
while lower-energetic photons Compton-scatter off electrons. The correct treatment of energy injection within the 
molecular cloud in which the first star forms is thus critical.  Ideally, one would numerically compute the stages of energy injection at different radii throughout the star.

\section{Detectability of Dark Stars with the James Webb Space Telescope}
\label{sec:JWST}

In Ref.\cite{SMDS} and \cite{Ilie12}, some of the authors of this paper, as well as Ref.\cite{Zackrisson2010},
 studied the capability of 
 the upcoming James Webb Space Telescope (JWST) to
detect supermassive dark stars (SMDSs). As shown above, if  the first stars are powered
by dark matter heating in triaxial dark matter halos, some of them may grow to
be very large $> 10^6 M_\odot$ and very bright $> 10^9 L_\odot$. These
SMDSs would be visible in deep imaging with JWST and even the Hubble Space
Telescope (HST) \cite{Zackrisson2010b, SMDSHST}.  

The Hubble Space Telescope (HST) has carried out a series of deep imaging surveys (the Hubble Ultra-Deep Field (HUDF),
Extreme Deep Field (XDF)) of a small patch of sky in the constellation Fornax in order to detect galaxies 
in the early Universe using the near infrared camera WFC3 with multiple broad-band filters 
(centered on 1055.2 nm (Y-Band), 1248.6 nm (J-Band), and 1536.9 nm (H-band)).  HST has
successfully identified galaxies at redshifts
$z \sim 7-10$, using the dropout technique described below (see Ref.\cite{Bouwens2010Blue, Bouwensetal11, Oeschetal11}). 

JWST will be a large infrared telescope with a 6.5-meter primary mirror and is expected to 
be launched in 2018. JWST is an international collaboration between NASA, the European Space Agency (ESA), and 
the Canadian Space Agency (CSA). Four infrared instruments are being built for JWST: the Near InfraRed Camera (NIRCam),
the Near InfraRed Spectrograph (NIRSpec), the Mid-InfraRed Instrument (MIRI), and the Fine Guidance Sensor/ Near InfraRed Imager
and Slitless Spectrograph (FGS-NIRISS) and are sensitive to wavelengths from 600 nm (optical red) to 28,000 nm.

\begin{figure*}
\begin{center}
\includegraphics[scale=0.25]{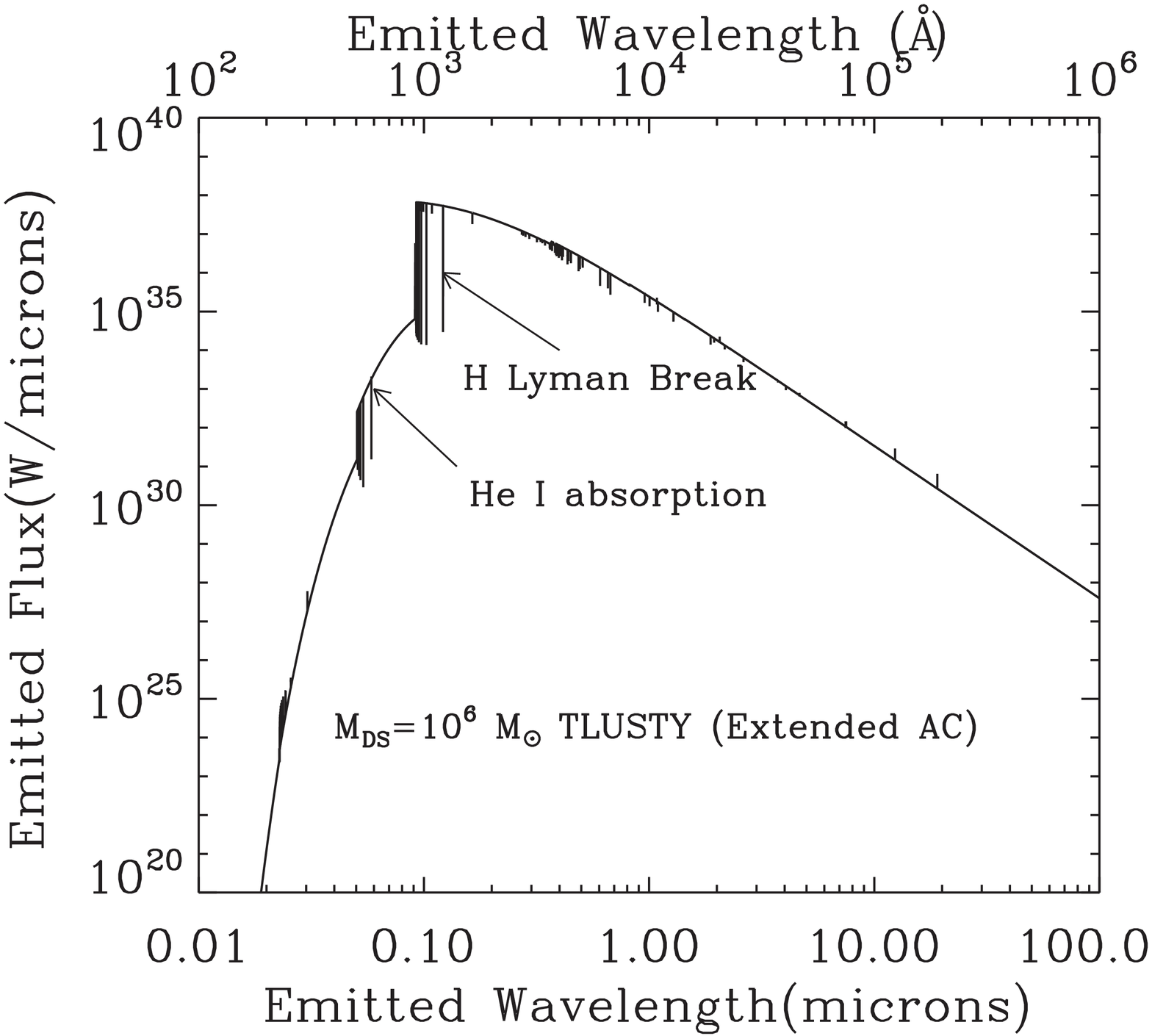}
\includegraphics[scale=0.25]{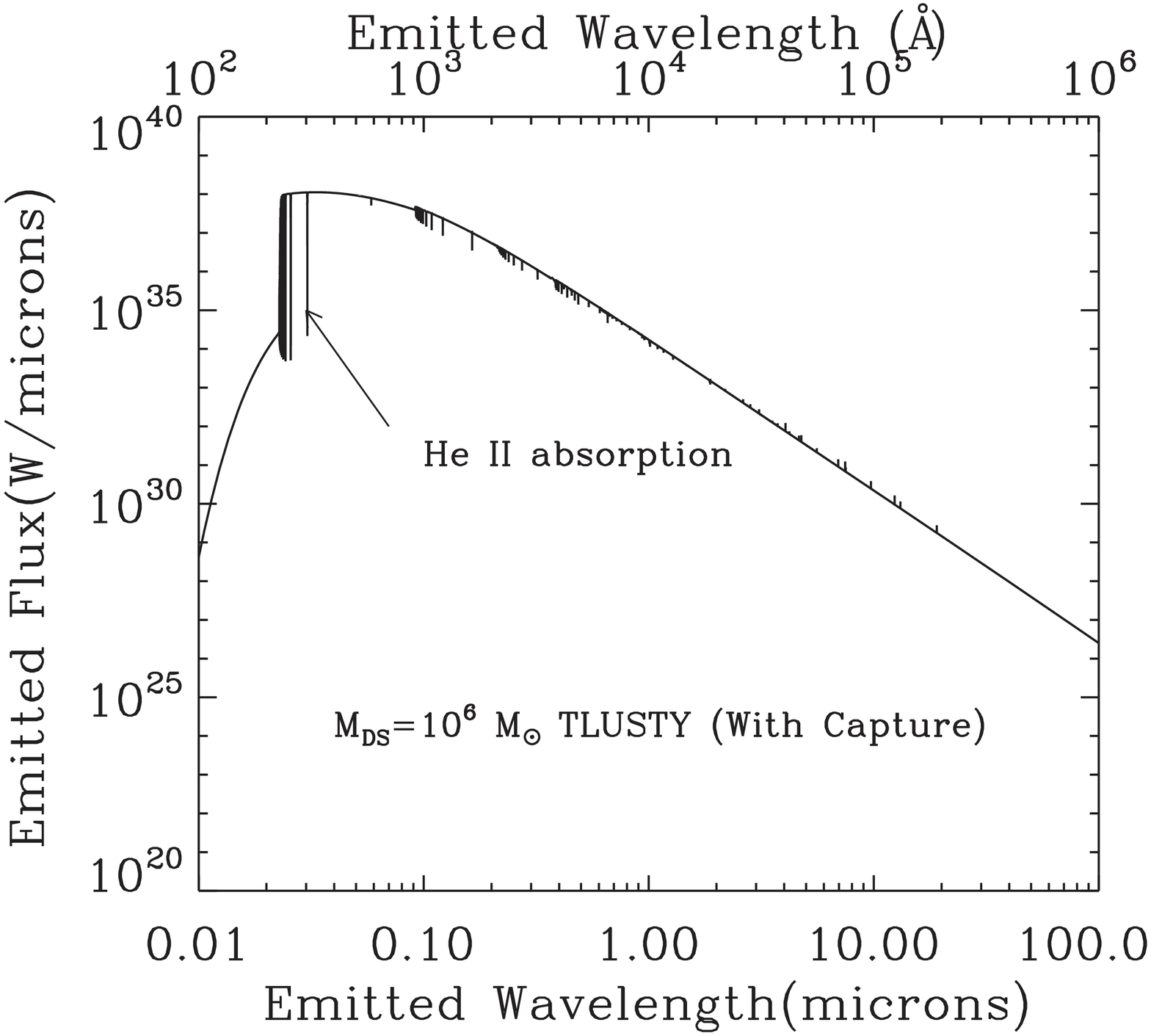} 
\end{center}
\caption{Expected spectral energy distribution (SEDs) of $10^{6} M_\odot$ supermassive DSs. Left panel: 
DS with a surface temperature of $1.9\times 10^4$ K and formed via the extended adiabatic contraction (AC) mechanism only. 
Right panel: with a surface temperature of $5.1\times 10^4$ K, formed via AC and ``with capture'' 
(Figure taken from Ref.\cite{Ilie12}).}
\label{fig:TLUSTYRest}
\end{figure*}

In order to determine the sensitivity of the HST and JWST instruments to DSs, some of us, with the help of Pat Scott, found
what SMDSs would look like \cite{Ilie12}, using the atmospheric models from Ref.\cite{Zackrisson2010b}. The spectra of SMDSs were obtained, 
using the publicly available TLUSTY \cite{TLUSTY} synthetic stellar 
atmospheres code. This code accounts 
for not only the black body radiation from the photosphere of the DS, but also accounts for absorption 
lines arising from the cooler gas in the atmosphere of the star.  The spectra for the case of a $10^6 \msun$ DS that formed via the two mechanisms of extended AC and capture  are shown in Figure~\ref{fig:TLUSTYRest}.

\begin{figure*}
\begin{center}
\includegraphics[scale=0.25]{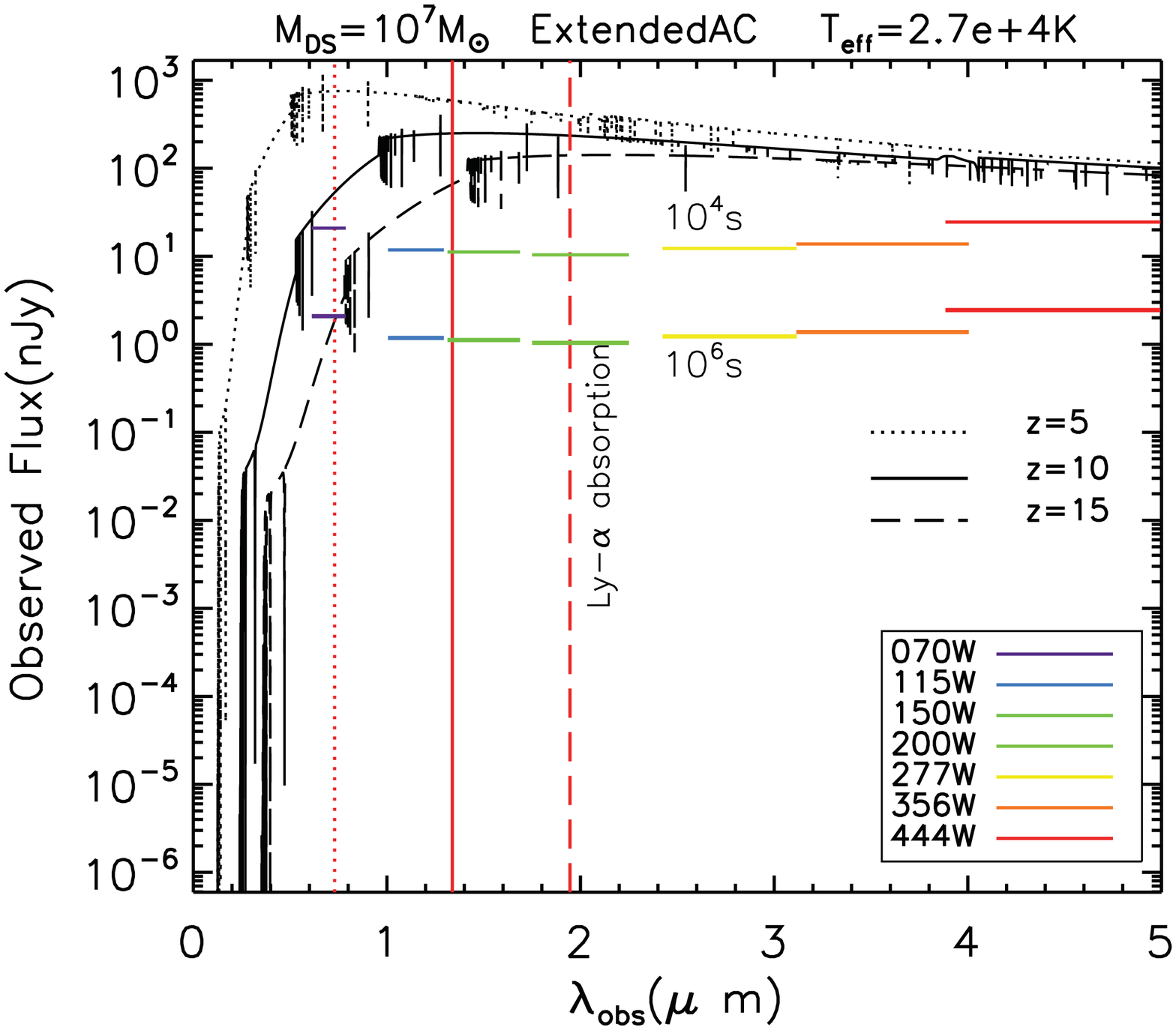}
\includegraphics[scale=0.25]{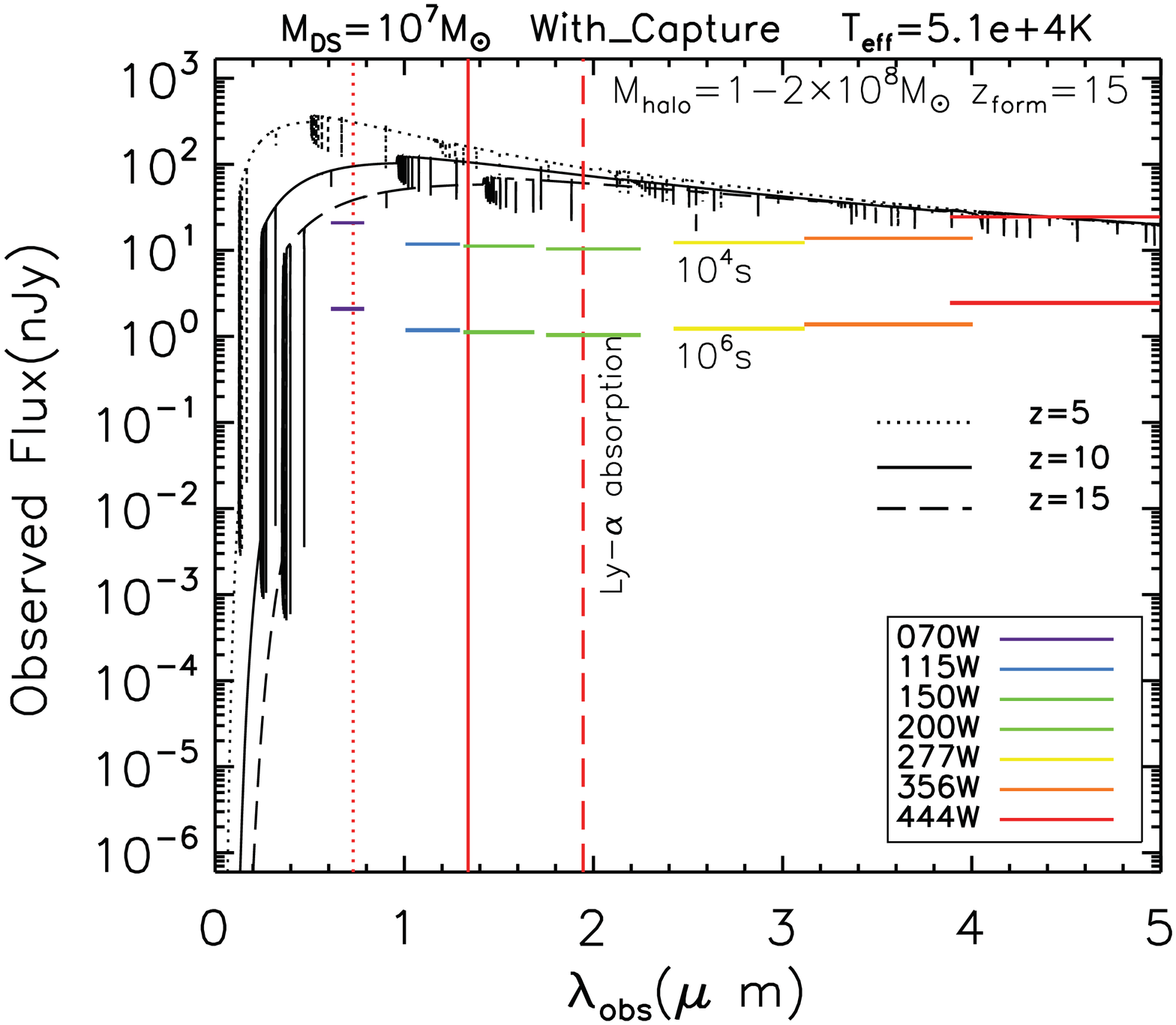}\\
\vspace{0.3cm}
\includegraphics[scale=0.25]{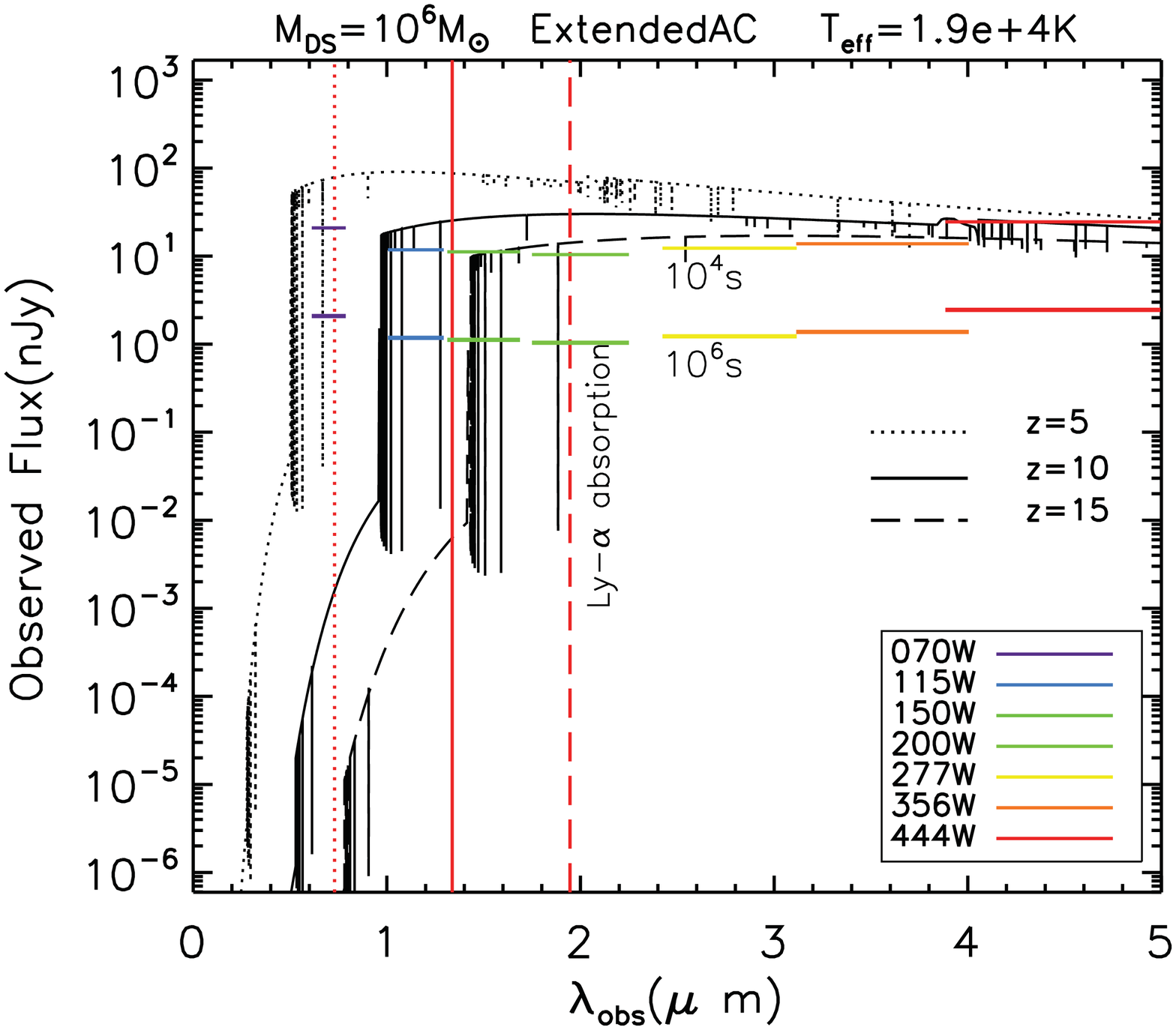}
\includegraphics[scale=0.25]{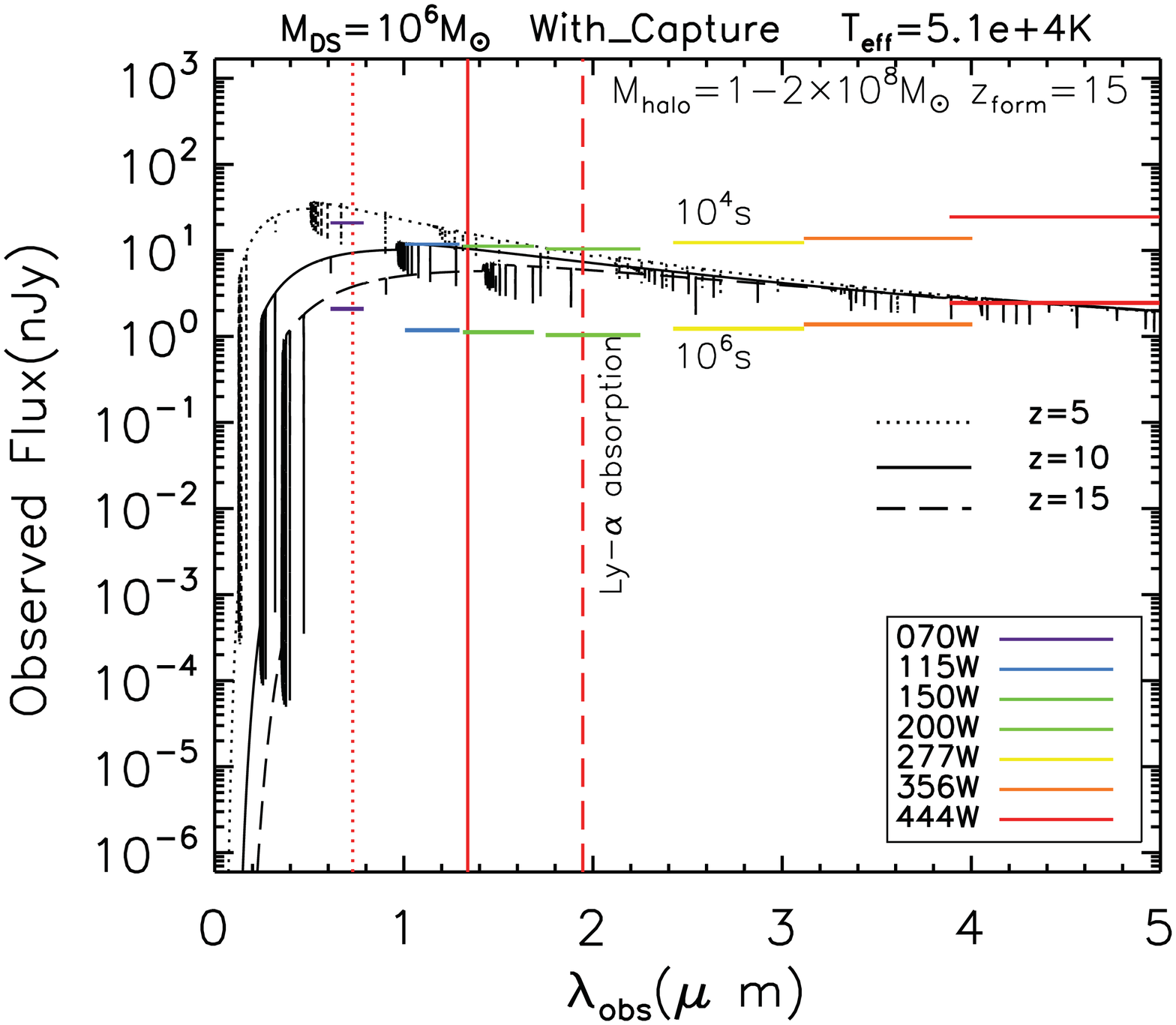}\\
\end{center}
\caption{Spectra for supermassive DSs formed at $z_{\rm form}=15$ (formation redshift) compared with sensitivity of JWST filters. 
Listed above each panel are the mass of the DS in solar masses, the formation mechanism (extended AC or ``with  capture'')
and the surface temperature $T_{\rm eff}$. The fluxes are shown at $z=15$ (dashed line), $10$ (solid line) and $5$ (dotted line) and compared to the detection limits of NirCam wide passband filters. The colored horizontal lines represent the sensitivity limits for the filters as labeled in the legend for exposure times
$10^4$ sec (upper lines) and  $10^6$ sec (lower lines).  IGM absorption will decrease the observed fluxes for wavelengths 
shortward of the vertical red lines, which indicate the Lyman-$\alpha$ 
line (1216 \AA) redshifted  from the rest-frame of the star (Figure taken from Ref.\cite{Ilie12}).}
\label{fig:TLUSTYRedshifted}
\end{figure*}

We studied the sensitivity of JWST to detect dark stars in \cite{Ilie12}.  We showed that SMDSs in the mass range
$10^6-10^7 M_\odot$ are bright enough to be detected in all the wavelength bands of the NIRCam on JWST 
(but not in the less sensitive MIRI camera at higher wavelengths) with ultra deep exposures of $10^6$ seconds. Figure~\ref{fig:TLUSTYRedshifted} 
illustrates the spectra for SMDSs and compares them to the sensitivities of JWST filters.

Also, we used sensitivity limits from previous HST surveys to place bounds on the numbers of SMDSs that may be detected in future
JWST imaging surveys, following the approach of Ref.\cite{Zackrisson2010b}. Indeed, \cite{Zackrisson2010b} and  \cite{Ilie12} 
showed that SMDSs of $10^7~\msun$ are so bright they would have already been detected by the 
HST extreme ultra deep field surveys, if they existed at redshifts of 12 or lower. The absence of a DS detection in the
HST deep field surveys probably rules out supermassive DSs of this mass, but less massive DSs (e.g. $\lesssim 10^6~\msun$)
could exist. Also, the probability of detection depends on the unknown fraction of dark matter halos that could host a dark star. 

The ``dropout" technique can be used to determine the redshift of an object that is too faint for spectroscopic study.
Here the object is detected in filters centered at one (or more) frequencies 
but not in lower frequency filters;
it is invisible in the lower frequency bands because the light has been swallowed by Lyman-$\alpha$
absorption.  The redshift of the object can then be identified, because of the knowledge of the
frequency at which the Lyman-$\alpha$ absorption was significant.  The objects will be detectable in filters centered at higher wavelengths but not in the band 
 including the redshift of the object due to Lyman-$\alpha$ absorption.  Objects existing
at $z \sim$10, 12, or 14 will be detectable as J-band, H-band, or K-band dropouts, 
 respectively

\begin{figure*}
\begin{center}$
\begin{array}{cc}
\includegraphics[scale=0.25]{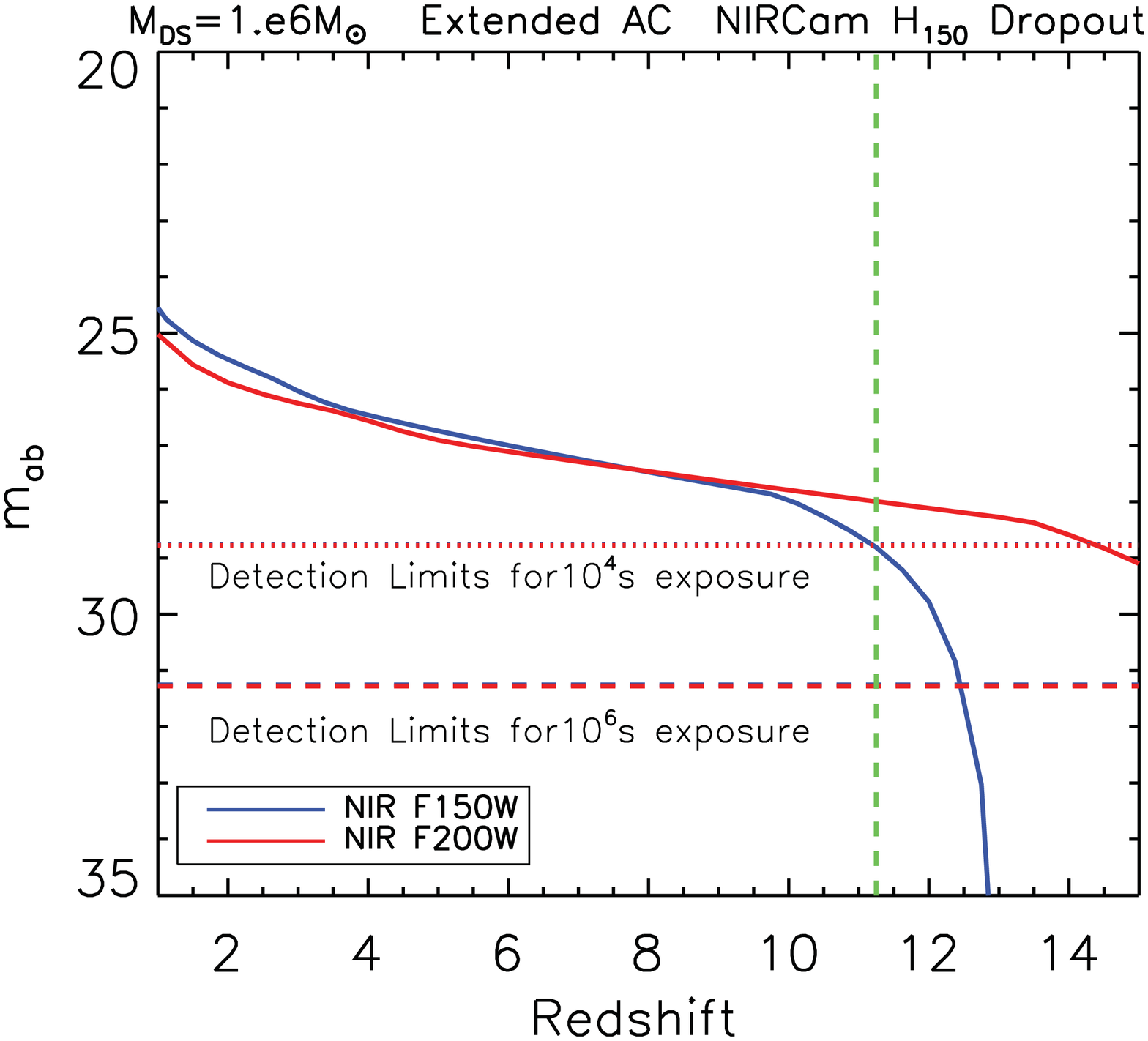} &
\includegraphics[scale=0.25]{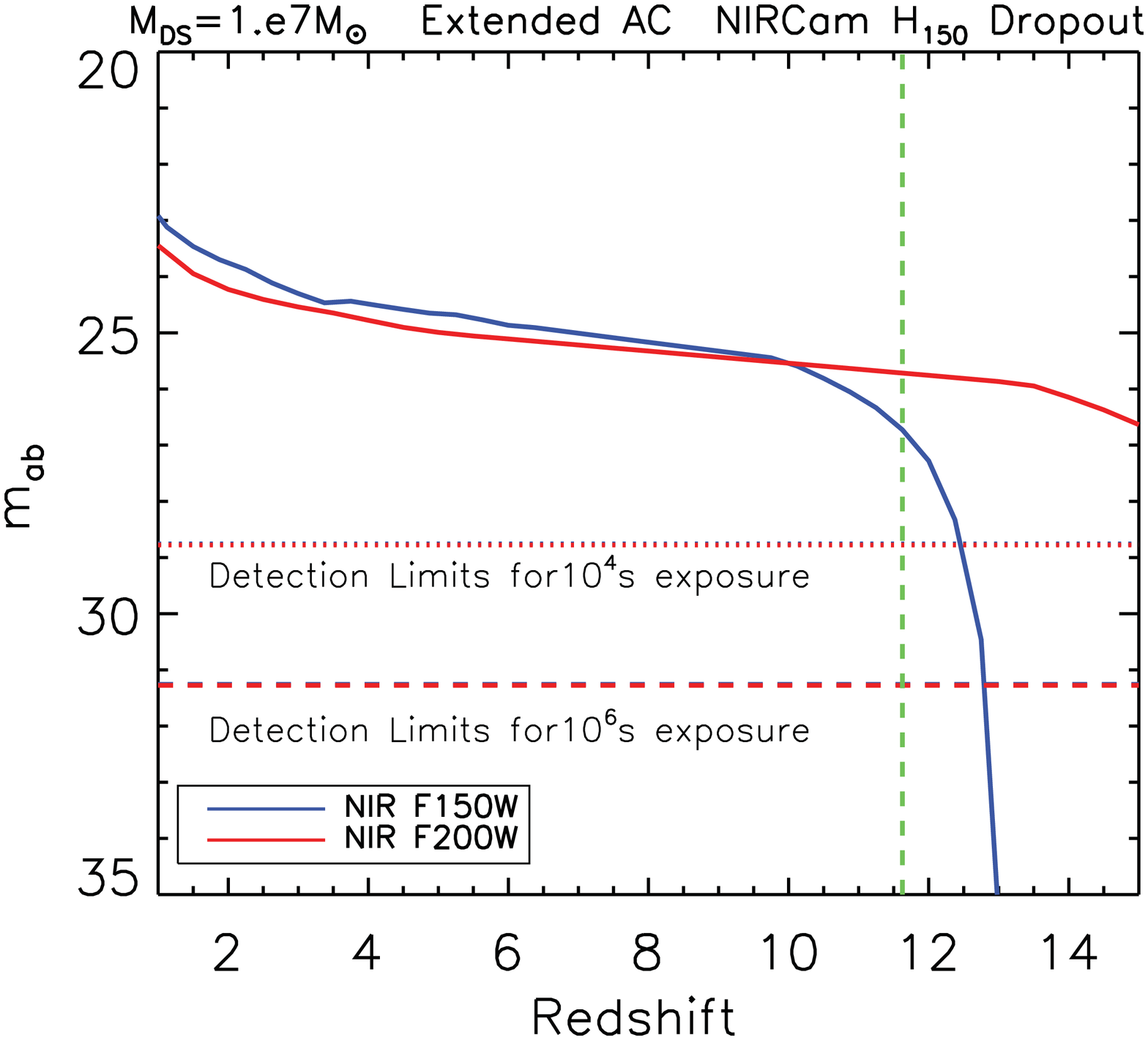} \\
\includegraphics[scale=0.25]{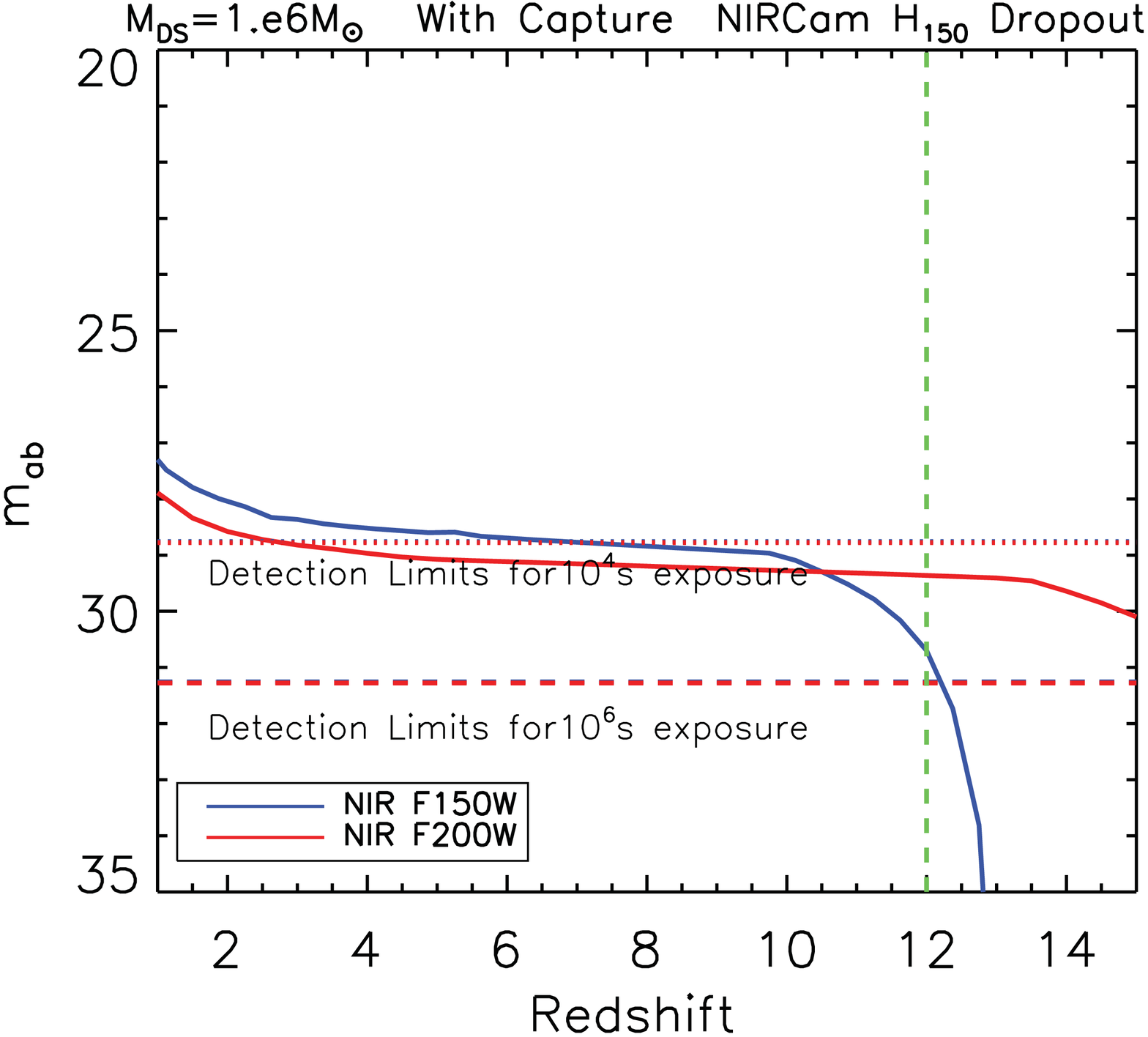} &
\includegraphics[scale=0.25]{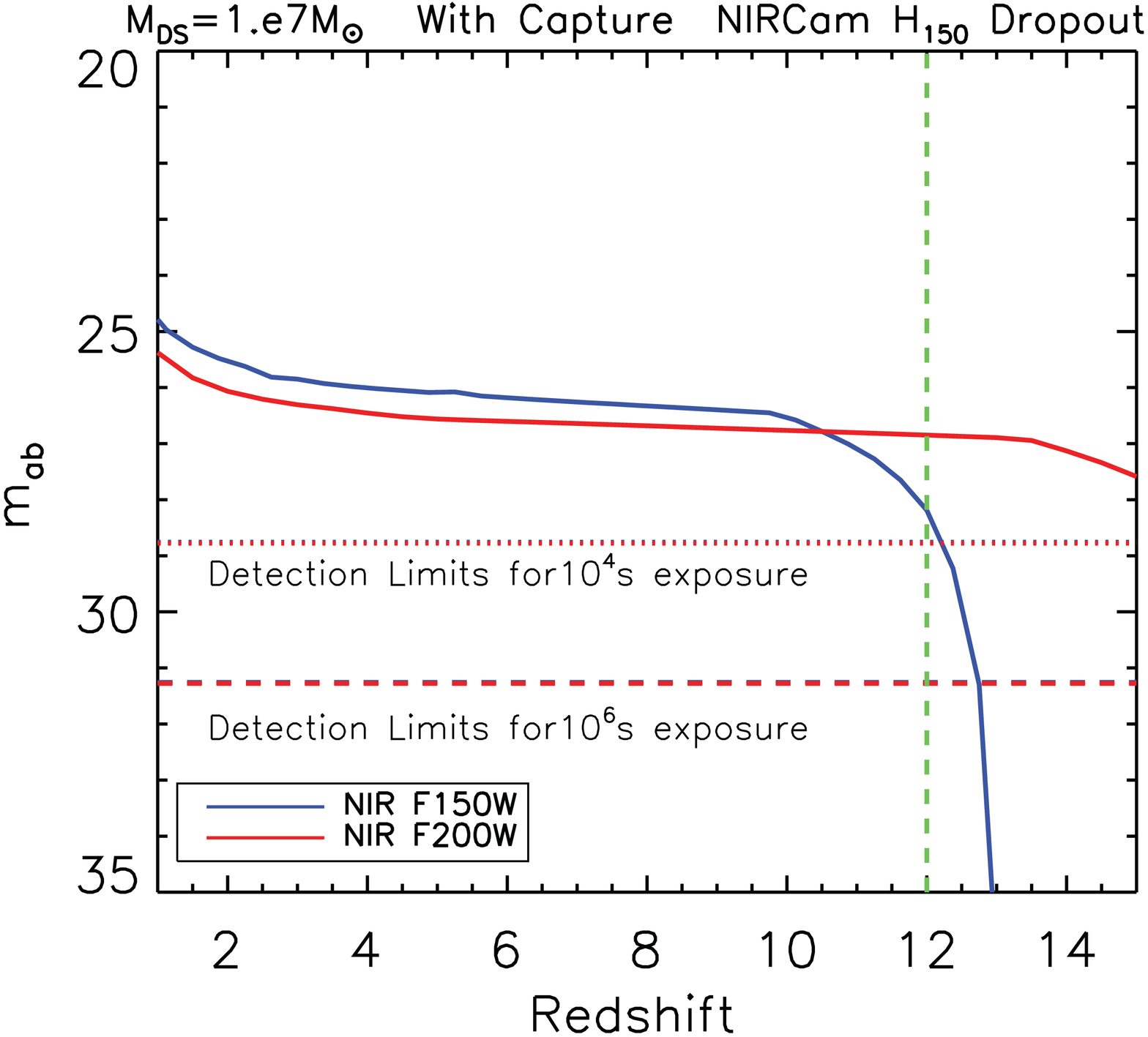} \\
\end{array}$
\end{center}
\caption{Supermassive dark stars with JWST as \NirHband band dropouts: Apparent magnitudes for SMDSs through the F150W 
(central wavelength 1505 nm) and F200W (central wavelength 2000 nm) NirCam filters. Those could be used to 
establish dropout detection criteria in the $12-14$ redshift range.  Top panel: 
$10^6\msun$ and $10^7\msun$ DSs formed without considering DM capture. Lower panel: $10^6\msun$
and $10^7\msun$ DSs formed including DM capture. The vertical green dashed line indicates the minimum 
redshift at which the DS will appear as a dropout (Figure taken from Ref.\cite{Ilie12}), where also illustrates
detectability of DSs in other
wavelength bands than the ones discussed here.}
\label{Nir150200}
\end{figure*}

HST has been able to find objects as J-band dropouts, meaning that their redshifts are $z \sim 10$.
 Whereas JWST is not particularly better than HST at finding J-band dropouts, it will be
significantly better at finding SMDSs as H-band and K-band dropouts at higher wavelengths,
corresponding to objects at $z \sim 12$ and 14, respectively. In the case of H-band dropouts,  
the object can be seen in the F200 NIRCam filter of JWST, but not in the F150 NIRCam filter.  
We required the difference between the broadband fluxes in the \NirHband and \NirKband filters to be greater than 1.2 AB magnitudes.
We can see that the SMDSs stellar light seen with JWST's \NirKband filter is essentially unaffected by  
Lyman$-\alpha$ absorption until $z\sim 15$, whereas  the IGM absorption will cut off most of the flux in 
the \NirHband at $z\gtrsim 11.5$ (see Figure~\ref{Nir150200}).   Hence, SMDSs can appear as H-band dropouts. 

 We estimated the number of SMDSs that JWST should be able to discover.
  With a total survey area of 150 arcmin$^2$ 
 (assuming a multi-year deep parallel survey), we found that typically the number of $10^6 M_\odot$ SMDSs 
expected to be discovered as H or K-band dropouts  is $\sim 10^5 f$, where the fraction of early DM halos hosting a dark star
is likely to be small, $f \lesssim 1$.  If SDMSs survive down to $z=$10 where HST bounds apply, then the observable number of SMDSs as H or K-band dropouts with JWST is $\sim 1-30$.

While individual SMDSs are bright enough to be detected by JWST,
standard Pop III stars (without dark matter annihilation) are not; 
they would instead contribute to the detectability of first galaxies with total stellar masses of
$10^6-10^8 M_\odot$. It will be interesting to re-assess DS detectability in more detail, in
light of the new MESA results. 
Differentiating first galaxies at $z > 10$ from
SMDSs may be possible with spectroscopy.  As described in the next section, a new interesting method of differentiating
SMDS from galaxies will take advantage of the fact that SMDSs may pulsate, with timescales
of the order of months in the observer's frame.  If cool, bright objects are found to
pulsate, then they are likely SMDSs.

\section{Dark star pulsations: a new avenue of detectability}
\label{sec:pulsations}

\begin{figure}
\centering
\includegraphics[scale=0.45]{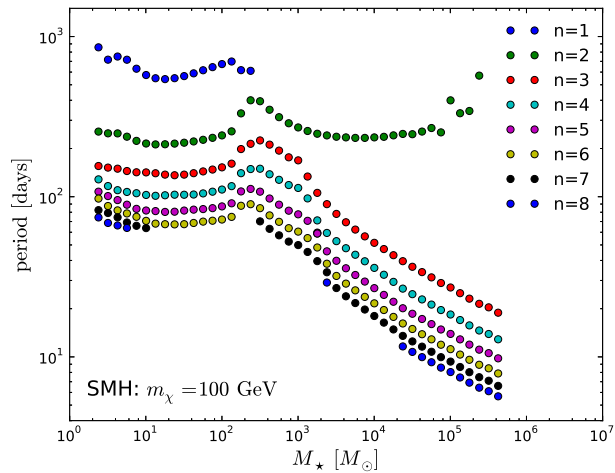}\\
\hspace{0.5cm}
\caption{Radial, adiabatic pulsation periods as a function of DS mass for a WIMP mass of 100 GeV and a DS forming in SMH.
The periods are given in the restframe of the DS.
The curves are for different overtone number, from the fundamental radial oscillation $n=1$ (upper-most curve) to 
$n=8$ (lower-most curve); see also Ref.\cite{TRDetal14}.}
\label{fig:periods}
\end{figure}

Some of the authors of this review have embarked on the study of an entirely new avenue of DS astrophysics, 
namely DS oscillations and pulsations. 
The usage of a fully-fledged stellar evolution code like MESA permits us to study such pulsations, which are deviations from hydrostatic equilibrium, 
see Ref.\cite{TRDetal14}. There are traditionally two classes of stellar pulsations: acoustic modes, or p-modes,
where pressure is the restoring force, and gravity modes, or g-modes, where buoyancy, i.e. gravity, is the restoring force
(the limit case of 'surface gravity modes' are also called f-modes).
Our analysis shows that g-modes are most likely not to be found in supermassive DSs, since their interior, albeit 
radiation-dominated, is subject to weak convective instability; however, p-modes are permitted. As a first step, we calculated
the adiabatic pulsation periods of radial modes (i.e. those for which there is no angular dependence, so $l=0$) with different
overtone number $n$, where $n=1$ is the fundamental or ``breathing`` mode, and $n>1$ are higher overtone modes. 
In Figure \ref{fig:periods}, we plot the restframe pulsation periods as a function of DS mass for the halo environment SMH in 
eq.(\ref{eq:defineSMH}) and a WIMP mass of 100 GeV. The periods are considerably shorter for higher overtone numbers.
For the example shown in Figure \ref{fig:periods}, the restframe of a DS with $10^5 M_{\odot}$ is at $z \simeq 14.82$. Converting to the
observer's frame, the periods range from 6323 days for $n=2$, down to 127 days for $n=8$. 

In general, we find that our DS models pulsate on timescales which range from less than a day to more than two years
in their restframes at about $z=15$, depending on the WIMP mass and overtone number. The pulsation periods are significantly
shorter for higher WIMP mass. Converting to the observer's frame, the shortest periods we found are less than about 50 days for
modes with $n>6$ and a WIMP mass of 1 TeV \cite{TRDetal14}.

Work is in progress to study DS pulsations more quantitatively, in particular possible driving mechanisms of DS pulsations.
Preliminary results suggest that the traditional $\kappa$-mechanism could occur in DSs,
as well. Here, pulsations are driven in layers of (partially) ionized hydrogen and helium where opacities are
high. Another possible driving
mechanism may be related to DM itself: small perturbations of the DSs could lead to local changes in the baryonic and DM densities,
modulating the DM heating rate, in turn\footnote{This would be akin to the $\epsilon$-mechanism in normal stars, where fusion in the stellar
cores can be subject to variation due to local temperature changes, resulting in stellar pulsations.}.
If SMDSs are found to pulsate, this would represent yet another way of distinguishing DSs observationally 
from galaxies at high redshifts. If the pulsations are detectable, DSs may in principle someday be used as novel standard
candles for cosmological studies.

\section{Dark stars as seeds for supermassive black holes in the Universe}
\label{sec:afterlife}

Supermassive black holes (SMBHs) are ubiquitous throughout the Universe.
Observations of quasars at redshifts above $z \sim 6$ suggest the presence of 
SMBHs in excess of $10^9 M_{\odot}$ at a time when the Universe was not even a billion years old \cite{Fan:2001}, see
also \cite{Wu:2015} for a recent discovery of a SMBH of an estimated 12 billion $M_{\odot}$.
In addition, SMBHs
are found at the centers of nearly all galaxies, and their masses appear to be well correlated with numerous properties of
the spheroidal components of their host galaxies \cite{kormendy_ho_13}.
The origin of the scaling relations between host galaxies and their SMBHs remains an actively researched topic. 
One issue that remains unsolved is the  origin and mass distribution of the seed black holes that eventually grow 
into supermassive  ones. It is generally believed that there is not enough time for stellar mass black holes that 
result from the death of massive stars to accrete at a high enough rate to grow into the SMBHs that power high redshift quasars.
Some physical mechanism is required to produce intermediate-mass seed black holes of $10^3-10^5 M_{\odot}$ onto which further
accretion can take place. One picture of black hole seed formation is the direct collapse of gas clouds in the 
center of minihalos, see \cite{1994ApJ...432...52L}. Begelman's~\cite{2010MNRAS.402..673B} ``quasi-stars'' offer 
another route to the formation of SMBHs.
Another new idea for explaining the existence of SMBHs has been proposed in~\cite{2014Sci...345.1330A},
where a black hole moving inside of a star cluster
can have rapid accretion and become very massive.

Recently, subsequent to our work on supermassive dark stars, there has been renewed interest in supermassive
stars (SMSs) of purely gas origin. These resemble SMDSs in many ways and may also serve as explanations of supermassive
black holes;
see Ref.\cite{Reisswig:2013sqa, 2012ApJ...749...37M, Chen:2014yea, 2013ApJ...778...17W, 2013ApJ...778..178H}
for more details. The proposed SMSs, however, require very high accretion rates, $\dot M \gtrsim 10^{-1} M_{\odot}$, and hence
can only form in atomic-cooling halos like LMH in Eq.(\ref{eq:LMH}). A detailed comparison of the structure and evolution between SMSs and supermassive dark stars
will be published elsewhere \cite{RDF}.
 
As we have seen, SMDSs could form in a variety of halo
environments, ending up with different final stellar masses.  
We described earlier in this review the case of SMH (see Eq.(8)).
Alternatively, in the ``Large Minihalo Case'' (LMH) of Eq. (9), a
hydrogen/helium molecular cloud may start to contract in a $10^8
\msun$ halo. Here 
the virial temperature is sufficiently high for the gas cloud to cool by hydrogen line cooling; then
the higher temperature implies a larger accretion rate of $10^{-1} ~M_{\odot}/\rm{yr}$.
We studied the growth of SMDSs using both polytropes and MESA for this LMH case. 
The amount of
baryonic mass in these larger halos is $1.5 \times 10^7 \msun$, and potentially, all
of this mass could go into the SMDS.  

Dark stars could provide seeds for the many supermassive BHs seen throughout the Universe
and at early times.
 Once the dark matter fuel inside a dark star is exhausted, 
then the star contracts to maintain pressure support.  Lighter DSs become fusion-powered for
$\sim 10^6$ years before collapsing to black holes. On the other end of the spectrum, the most massive supermassive dark stars may collapse directly to supermassive black holes with masses of $> 10^5 M_{\odot}$ without any fusion phase at all, similar to the monolithic
collapse of central gas clouds in the more traditional scenarios of forming supermassive black holes.
Either mechanism  would clearly help to provide seed BHs that might
explain the many SMBHs found throughout the Universe.

\subsection{Gamma-Ray Constraints}

$\gamma$-rays can constrain the mass and abundance of DSs. A series of papers \cite{Sandick:2011zs,Sandick:2010yd}
explored the detection prospects for $\gamma$-rays produced in dark matter annihilations in the DM spikes surrounding black holes 
for a range of star formation scenarios, black hole masses, and dark matter annihilation modes.   
Here we review the constraints obtained
by comparing the $\gamma$-rays measured by the FERMI Gamma-Ray Space Telescope ~\cite{Ackermann:2013yva}
with the predictions of $\gamma$-ray annihilation products from the remnants of early DSs in our Galaxy 
today (see also Ref.\cite{Bertone:2005xz}).
In hierarchical structure formation, smaller halos are incorporated
into larger halos. Some of the same halos which hosted the first stars now have merged inside our own Milky Way Galaxy. 
As a consequence, there may be a large number of remnant BHs inside our Galaxy.
 
 After a DS dies, the remaining BH can still be surrounded by a DM reservoir in the form of a dark matter ``spike."  
Annihilation of the DM inside the DS spikes can produce a copious number of $\gamma$-rays due to the high DM density.
  These $\gamma$-rays can be detected by FERMI.

 \begin{figure}[h]
\centering
\includegraphics{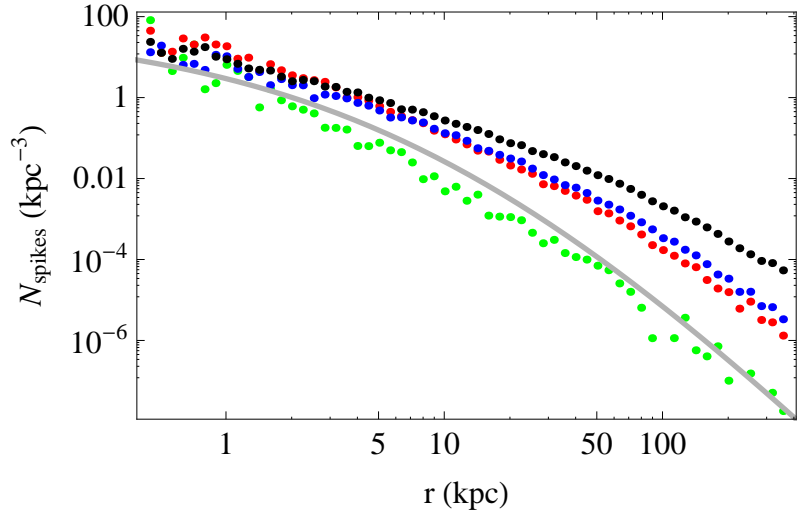}
\caption{The number density of black hole spikes in the Milky Way as a function of galactic radius for star formation models with Early (green), Intermediate (red), and Late (blue) $z_f$ as described in the text and assuming 100 percent efficiency in dark star formation.  Curves have been 
obtained using the Via Lactea (VL) II N-body simulation.  The black points illustrate the
total dark matter density profile at $z=0$ in VL-II; although the normalization of these points is arbitrary, it is useful 
to illustrate that the total DM profile is more extended than the distribution of black holes with DM spikes. 
Also shown as a solid grey curve is the analytical fit found in Ref.~\cite{2005PhRvD..72j3517B,2009PhRvD..79d3521T}. 
Our simulations show 409, 7983, and 12416 DM spikes in the Milky Way for the Early, Intermediate, and Late scenarios, 
respectively (Figure taken from Ref.\cite{Sandick:2010yd}. Reproduced by permission of IOP Publishing.).}
\label{fig:spikedists}
\end{figure}

 \begin{figure}
\centering
\includegraphics{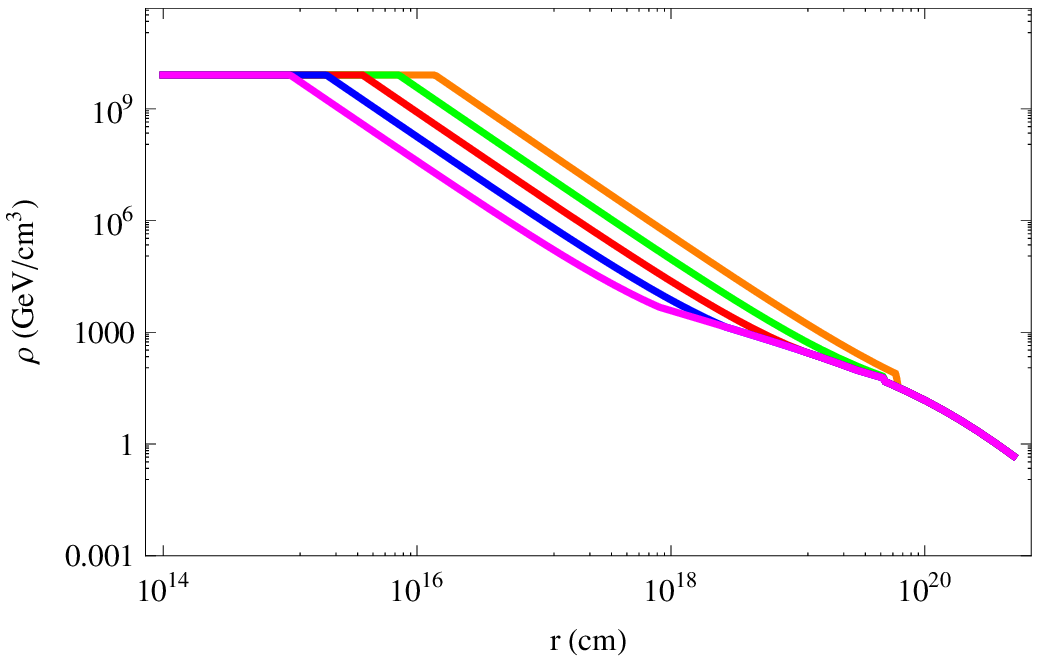}
\caption{Density profiles for contracted dark matter halos surrounding black holes of mass 
10 M$_\odot $(magenta), $10^2$ M$_\odot$ (blue), $10^3$ M$_\odot$ (red), $10^4$ M$_\odot$(green), 
and $10^5$ M$_\odot$ (orange), from bottom to top, assuming that the central black hole in each case 
formed at $z=15$ in a halo of mass $10^6 M_\odot$ (Figure taken from Ref.\cite{Sandick:2010yd}. Reproduced by permission of IOP Publishing.).}
\label{fig:densityprofs}
\end{figure}

Many black holes of mass $10 - 10^5 M_\odot$ that formed at the centers of minihalos survive in the Universe today. 
Assuming some fraction of high-redshift minihalos which hosted a DS, one can estimate the distribution of their remnant black holes today. 
Although some of the original minihalos would have merged with other DM halos, resulting in disruption, 
 one can still follow  the evolution of the black holes and their DM spikes in simulations. 
Some of us \cite{Sandick:2011zs,Sandick:2010yd} used
 the Via Lactea-II N-body simulation of Ref.\cite{2008Natur.454..735D} to track black hole spikes from the redshift of
 their formation to z = 0. 
In this way,  the black hole population could be estimated in a galaxy like our Milky Way today.

\begin{figure}
\begin{center}$
\begin{array}{cc}
\includegraphics[scale=0.45]{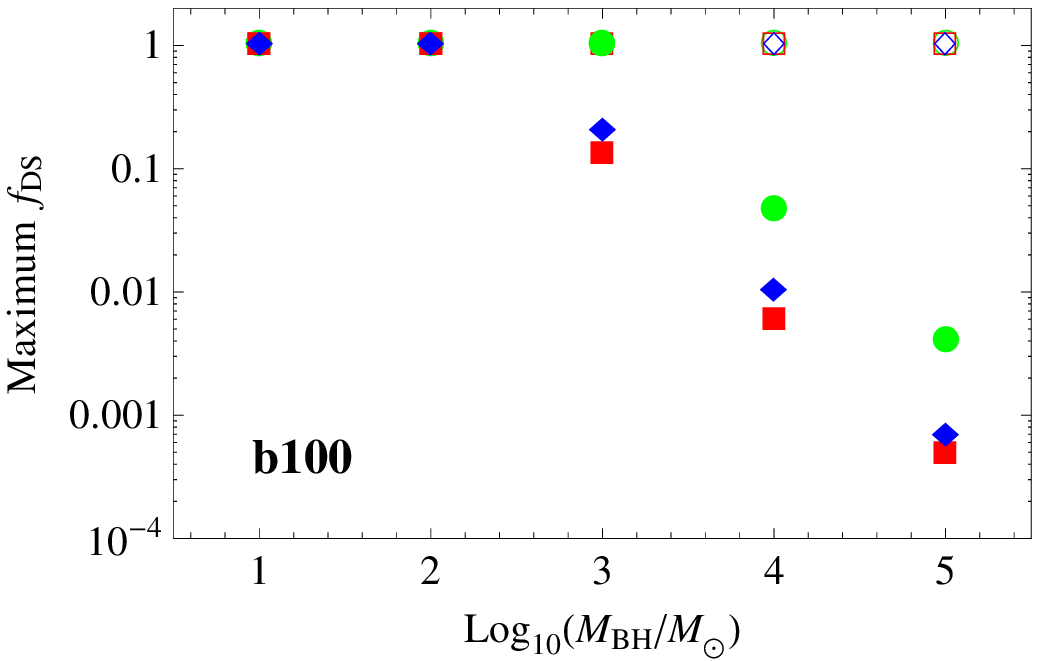}&
\includegraphics[scale=0.45]{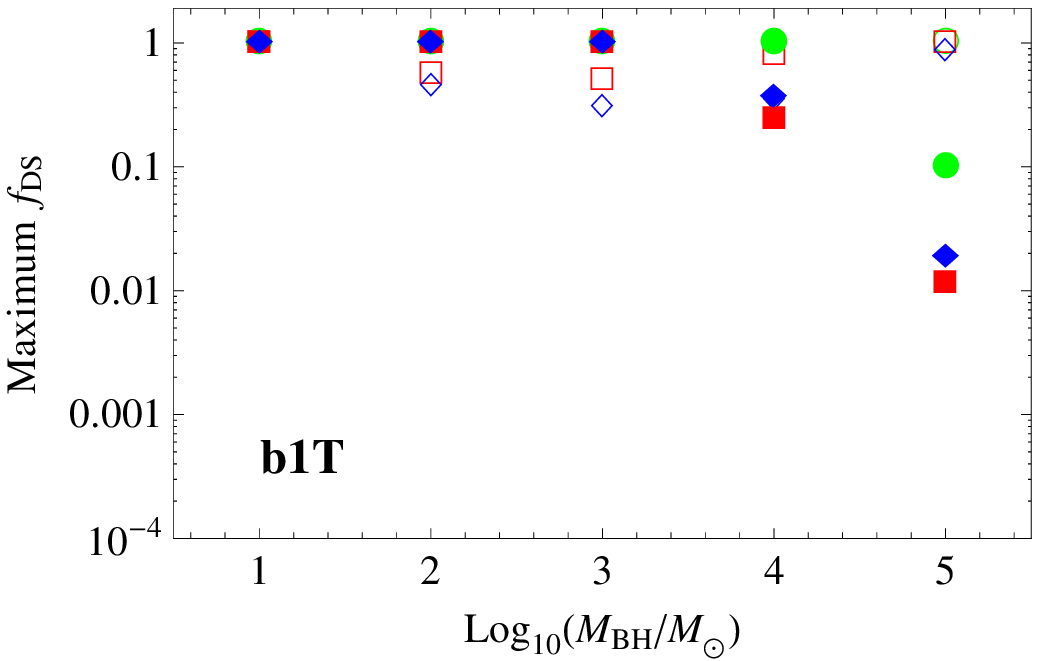}\\
\includegraphics[scale=0.45]{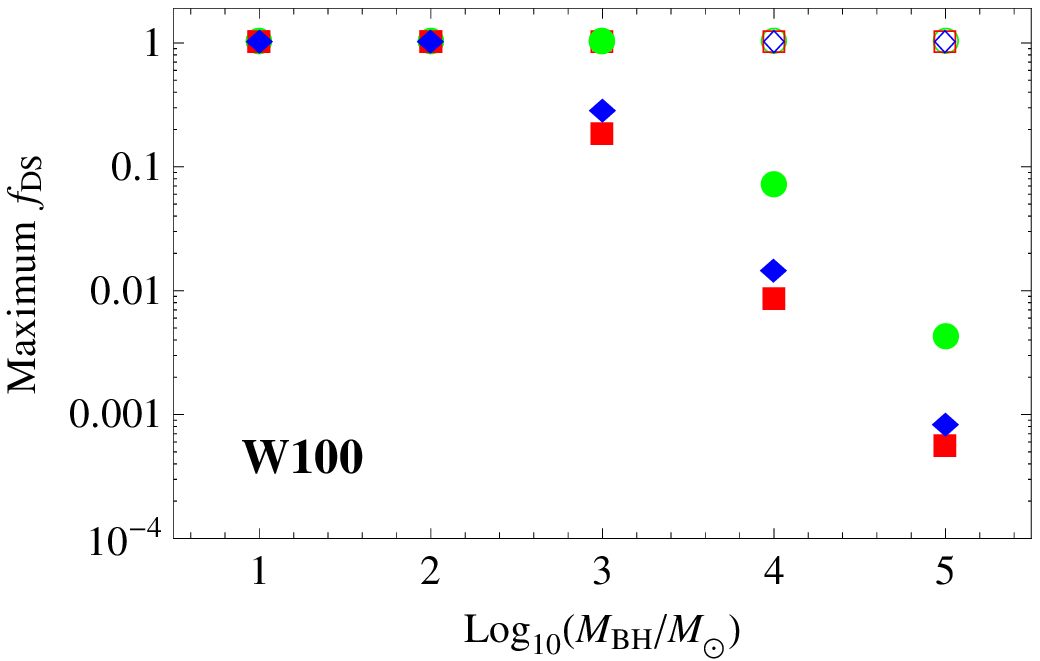}&
\includegraphics[scale=0.45]{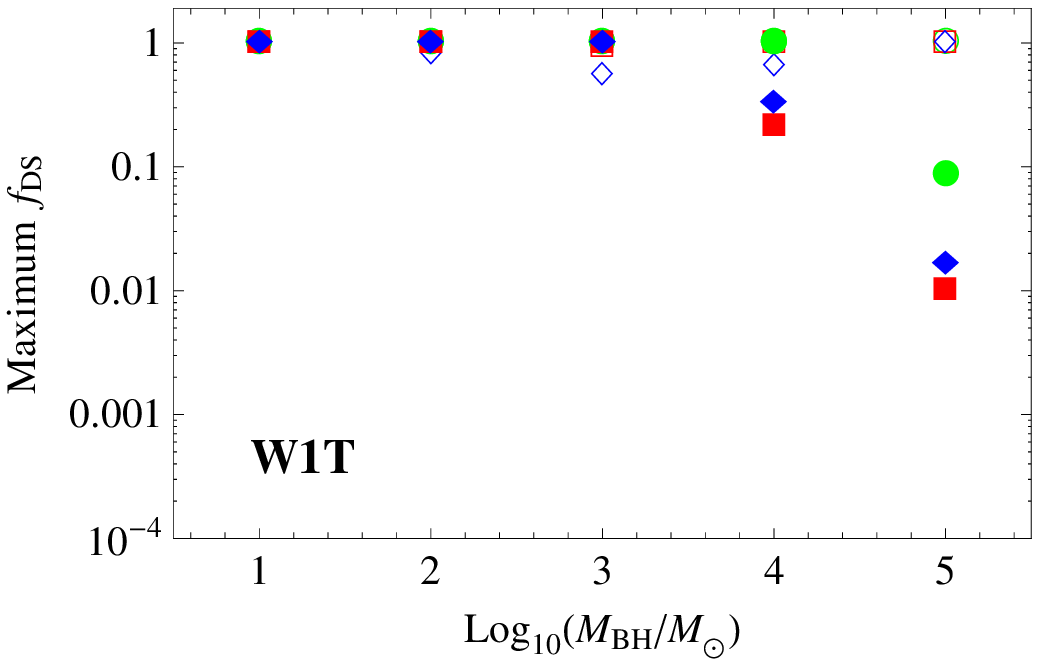}\\
\includegraphics[scale=0.45]{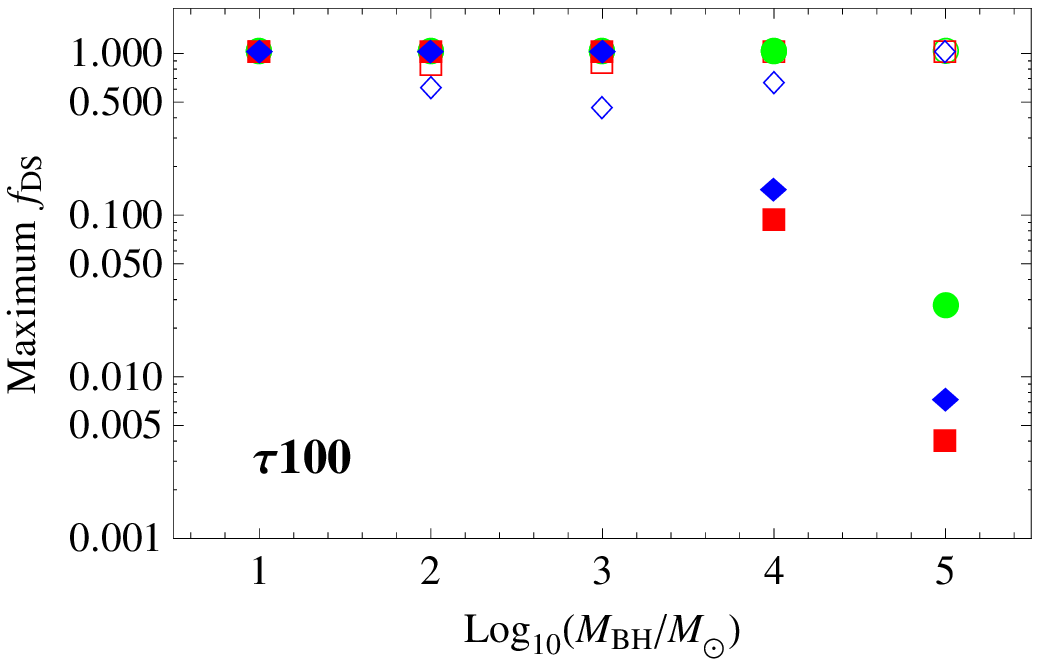}&
\includegraphics[scale=0.45]{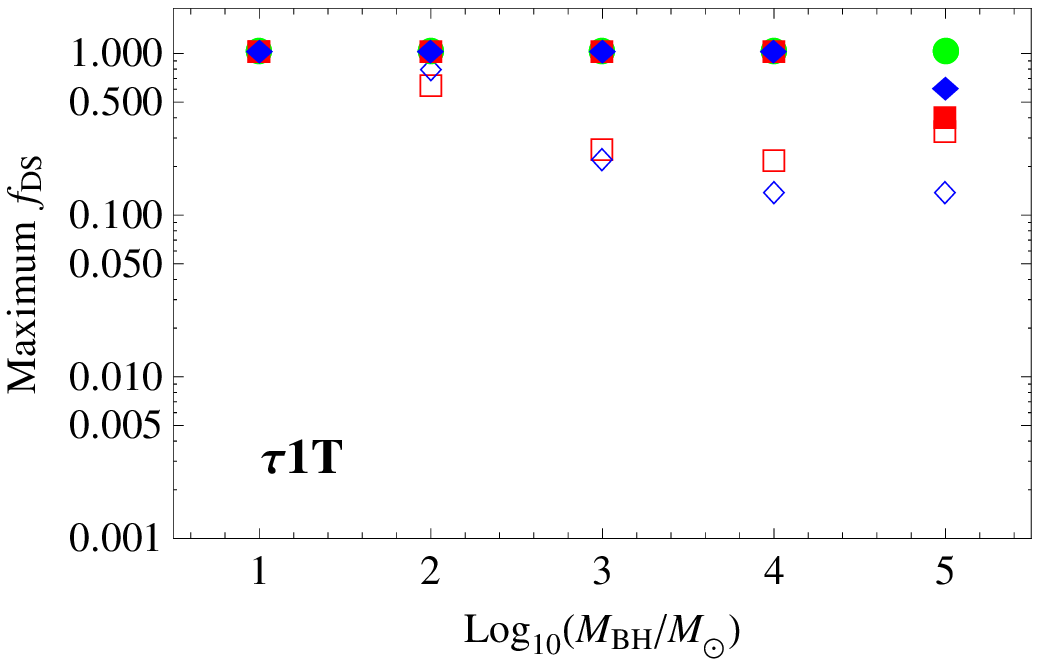}\\
\includegraphics[scale=0.45]{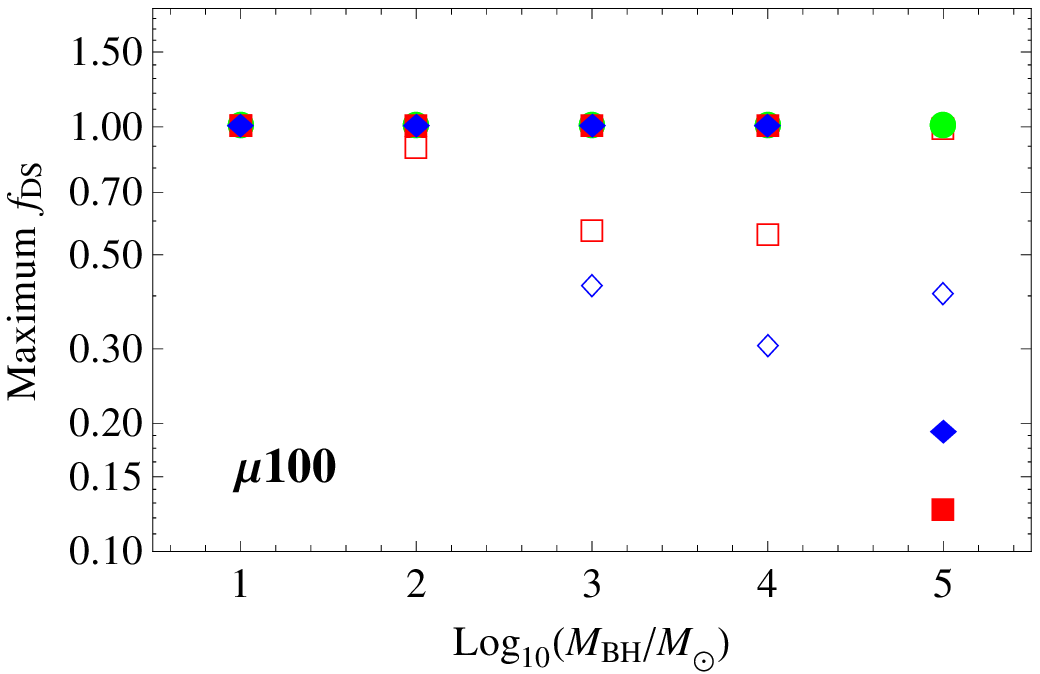}&
\includegraphics[scale=0.45]{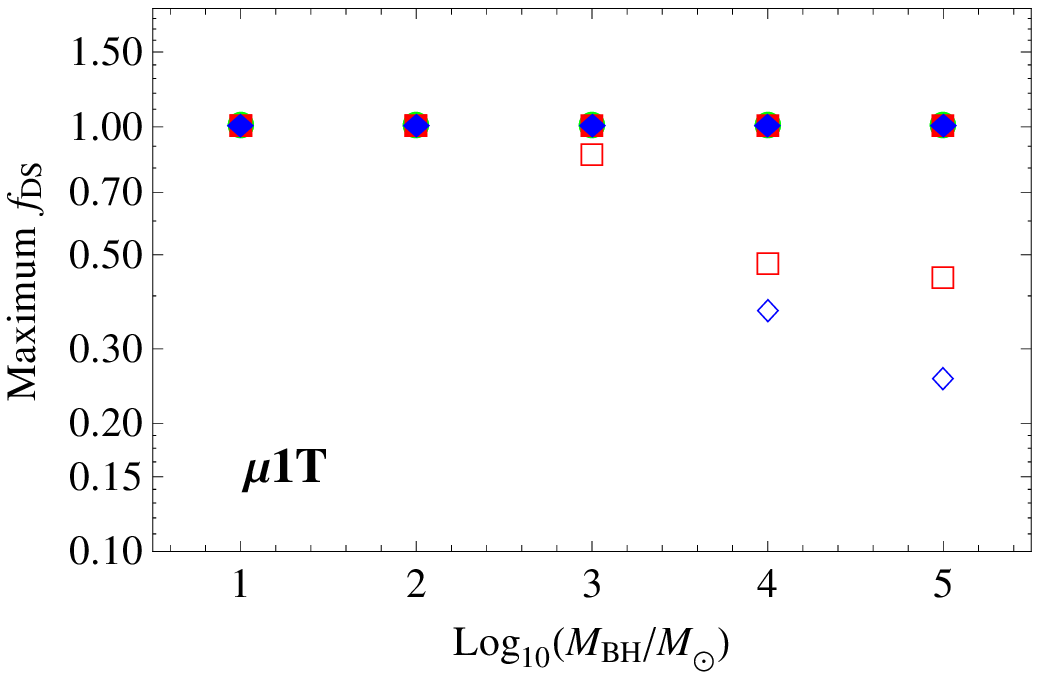}\\
\end{array}$
\end{center}
\caption{Maximum $f_{DS}$ (fraction of early minihalos which contain a DS)
as a function of central black hole mass for the WIMP annihilation into $\tau\bar{\tau}$, W$^+$W$^-$, $b\bar{b}$ and $\mu\bar{\mu}$.
 Green circles, red squares, and blue diamonds are for Early, Intermediate, and Late $z_f$, respectively. 
The left panels show $m_\chi=100$ GeV.  The right panels show $m_\chi=1$ TeV.
The solid markers are the limits from point source brightness, while the open markers are from the diffuse $\gamma$-ray flux. 
Each panel assumes a single annihilation channel. Note that the range of $f_{DS}$ displayed differs 
from panel to panel (Figure taken from Ref.\cite{Sandick:2010yd}. Reproduced by permission of IOP Publishing.).}
\label{fig:fDSmax}
\end{figure}

The number of spikes depends on an important parameter --- the range of redshifts during which the first stars 
can form.    
Since the redshift at which the increasing UV background and/or metal enrichment results in the truncation of first
star formation is poorly constrained, 
  three scenarios were examined for the end of first star formation: Early ($z_f = 23$), 
  Intermediate ($z_f = 15$), and Late ($z_f = 11$). 
  Figure \ref{fig:spikedists} illustrates the number density of BH spikes as a function of galactic radius 
  for these three cases.  The late scenario has the largest number of spikes.  
  
Figure \ref{fig:densityprofs} shows the contracted halo profiles today for DM spikes due to black holes of various masses 
for the case where the central object formed at z = 15. We note that the power law portion of the profile is independent of WIMP mass. 
In the central regions, closest to the black hole, some of the DM has annihilated away in the time since the formation of the central mass.
A more massive dark star will have more adiabatic contraction, which will
generate a larger spike, as seen in Figure \ref{fig:densityprofs}.
In addition, the DM particle properties will also affect the number of $\gamma$-rays produced.
For instance, low-mass DM will have a higher annihilation rate compared to a more massive DM particle (for a fixed cross section). 
Different  annihilation channels produce different
number of photons.  Leptons produce significantly fewer photons compared to gauge mediators and quarks.
Hence, a 10 GeV WIMP annihilating into $b\bar{b}$ will produce many more $\gamma$-rays compared to a 100 GeV WIMP annihilating
into leptons such as $\mu\bar{\mu}$.

The data from FERMI was used in a two-pronged approach to constrain the
 number of black holes in the Milky Way halo and, consequently, the number of DSs that could have formed at early times.
 First, the FERMI First Source Catalog \cite{FermiFirstCatalog} was used
 to find the minimal distance to the nearest DM spike such that it is not brighter than the brightest source observed by FERMI. 
 From the predicted distribution of such spikes in the Milky Way halo, a limit was extracted on the fraction of minihalos
 in the early Universe to host a black hole (and survive as a DM spike in our galactic halo today). Second, the FERMI measurement 
 of the diffuse $\gamma$-ray background was used to constrain the population of DM spikes contributing to the diffuse flux today, 
 thereby setting a second limit on the fraction of minihalos in the early Universe to become DM spikes in our Galactic halo.

These data were used to constrain the efficiency of DS formation $f_{DS}$,  
i.e., the fraction of early minihalos that contained a DS and their black hole remnants.
In Figure \ref{fig:fDSmax}, we show the constraints from point sources and from the diffuse $\gamma$-ray background.
One can see that the constraints are the strongest for very massive DSs and 
for annihilation into quarks and gauge bosons.
It is important to emphasize that the constraint is for halos which have a mass below $10^7$ M$_\odot$.
If the supermassive DSs form inside halos more massive than $10^7$ M$_\odot$, the constraints would be much weaker.

\section{Dark stars and their impact on reionization and radiation backgrounds}
\label{sec:backgrounds}

If DSs of various sizes constitute a part of early stellar populations, they will impact the history
of reionization, and affect the optical depth of the intergalactic medium as probed by the cosmic
microwave background. This allows to constrain the abundance and properties of DSs, albeit the analysis
is highly complicated by the fact that reionization models suffer from poorly known astrophysical parameters.
Only few studies have been done in the past so far to determine the effects of DSs on reionization,
and those were limited to DSs with masses smaller than about 1000 $M_{\odot}$. We stress that the ionizing photon flux for a
given stellar mass falls off very abruptly, as the relative contribution of DM annihilation heating to the star's total energy
budget is increased.
In order to study the impact on reionization, Ref.\cite{Schleicher:2009} 
considers a stellar population of either \textit{only} main-sequence dominated DSs (with some DM capture)
or DM-capture-dominated DSs. In neither population is DM heating dominant due to annihilation of 
adiabatically-contracted DM and DS masses are below about 1000 $M_{\odot}$.  
It is found that the former case delays reionization while the latter case hasten it. These results are in accordance
with the analysis of Ref.\cite{Scott:2011} in the respective limit cases. 
However, Ref.\cite{Scott:2011} considers more varied DS lifetimes and DS fractions of the total population
which does include Pop III and Pop II stars, as well. Furthermore, another population considered consists of DSs whose fuel
is dominated by adiabatically-contracted DM (and no capture), albeit only for a time as long as about 0.3-0.4 Myrs, at which point the DSs have
masses of about 800 $M_{\odot}$. This case corresponds to a DS history akin to the one depicted in Fig.\ref{fig:f2}. As we have pointed
out above, this case is pessimistic in terms of how long a supply of DM could be sustained. The 'dark star proper' considered in
Ref.\cite{Scott:2011} encompasses this case, as well as the transition of this case to phases of meager or extreme DM capture, in
which case the lifetime of the DS is extended up to 500 Myrs. We stress that this case does \textit{not} include those SMDSs which are 
predominantly fueled by the annihilation of adiabatically-contracted DM. DSs dominated by DM heating due to capture are always hotter and denser than those without capture included (see Section \ref{sec:stellarstructure} and Fig.\ref{fig:dens_f3}).
Therefore, their ionizing UV photon budget will be necessarily higher than that for the case without capture.

Ref.\cite{Scott:2011} find that, in the case where gravity is the only source of DM in the star  (i.e. DS without capture), DSs do not affect standard reionization. 
On the other hand, if dark stars capture large amounts of dark matter via nuclear scattering, Ref.\cite{Scott:2011} found that reionization can be substantially delayed,
leading to decreases in the integrated optical depth to last scattering and large-scale power in the EE polarization power spectrum. Thus, 
in principle observations of these quantities from CMB data could be used to bound 
DS stellar populations.  In the intermediate case, where moderate amounts of dark matter are captured by the DS, reionization can instead be sped up slightly, modestly increasing the CMB optical depth. 
Yet, as pointed out repeatedly in Ref.\cite{Scott:2011},  any effect of a DS population on reionization could be degenerate with the effects of the variation of (poorly known)
astrophysical parameters, particularly the star formation efficiency and the escape fraction of UV photons out of their halos.
Thus, it will be hard to ascertain whether DSs have a significant impact on the reionization history of the Universe.  Still, it may be possible to reconcile certain reionization models and their preferred astrophysical parameters with observations by
invoking the idea that some fraction of early stars consists of DSs. 

It would be important to re-do some of the previous investigations for the case of supermassive DSs 
(SMDSs) in the first stellar populations to see how reionization and the associated optical depth
may be changed.  As of yet, the reionization studies with DSs have only been done for DSs
weighing less than 1000 $M_\odot$.  This is a rich field which warrants more attention for future study.

Another probe of DSs is their impact on the extragalactic background light (EBL), whose main contribution comes from
integrated starlight and thermal dust emissions of all cosmic epochs since the
formation of the first stars. The EBL consists of the
optical to infrared part of the diffuse metagalactic radiation field, 
and has been identified as a
unique probe for the integrated star formation history of the Universe.
Ref.\cite{Maurer:2012} studied the effect of certain DS models on the EBL and noted
that, in comparison to other indirect DM detection signals, the DS-induced component in the EBL
is unique in that it is not sensitive to the exact branching ratios of the annihilation yields
into photons or charged particles and their resulting spectra, as all annihilation products
(except for neutrinos) are trapped and thermalized within the DS.
Model atmospheres were calculated in Ref.\cite{Maurer:2012} using the PHOENIX code for two DS models
from \cite{Spolyar09}, for DS mass $100-1000 M_\odot$, with effective surface temperatures between 5000-7500 K
(no significant hydrogen ionizing radiation is emitted in this range).  For comparison, we showed in Figure~\ref{fig:TLUSTYRest} the SMDS spectra obtained using the TLUSTY stellar atmospheres code for heavier stars
of mass $\sim 10^6 M_\odot$.  

The work of Ref.\cite{Maurer:2012} considered the constraints of the EBL on
DSs of 100-1000 $M_\odot$ only.
Since the DS formation rate and lifetimes are subject to high uncertainty,
a range of possible values was considered. Owing to the lower surface temperatures of DSs in this mass range,
the peak in the EBL is shifted towards higher wavelengths ($> 2 \mu$m), compared
to $\sim 1 \mu$m for standard Pop III stars of the same mass. By comparing lower and upper limits on the EBL,
the authors determined the allowed EBL contribution from DSs in the range of
5-25 nW m$^{-2}$ sr$^{-1}$ for wavelengths between 2 and 10 $\mu$m.
For example, a DS with $106~M_{\odot}$, $9\times 10^6 ~ L_{\odot}$, a lifetime
of $10^8$ yrs and a minimum formation redshift of $z=5$ results in a constraint on the DS formation rate between $5\times 10^{-4}$
and $3\times 10^{-3}$ yr$^{-1}$ Mpc$^{-3}$. 

It will be interesting to constrain more
DS parameters and evolution scenarios in the future, using their contribution to the EBL.  As shown in Figure \ref{fig:TLUSTYRest},
the peak wavelength for heavier SMDSs $\sim 10^6 M_\odot$ is lower than what was considered by Ref.\cite{Maurer:2012},
so that the constraints would be quite different.

\section{Dark Stars Existing Today}
\label{sec:today}

While it is possible for some of the first DM-powered stars to exist today, perhaps in under-dense regions of the Universe where
later star formation (with its accompanying ionizing photons) has not yet taken place, a more likely scenario is that later generations
of stars may also become DM-powered.

\begin{figure}
\begin{center}$
\begin{array}{cc}
\includegraphics[scale=0.25]{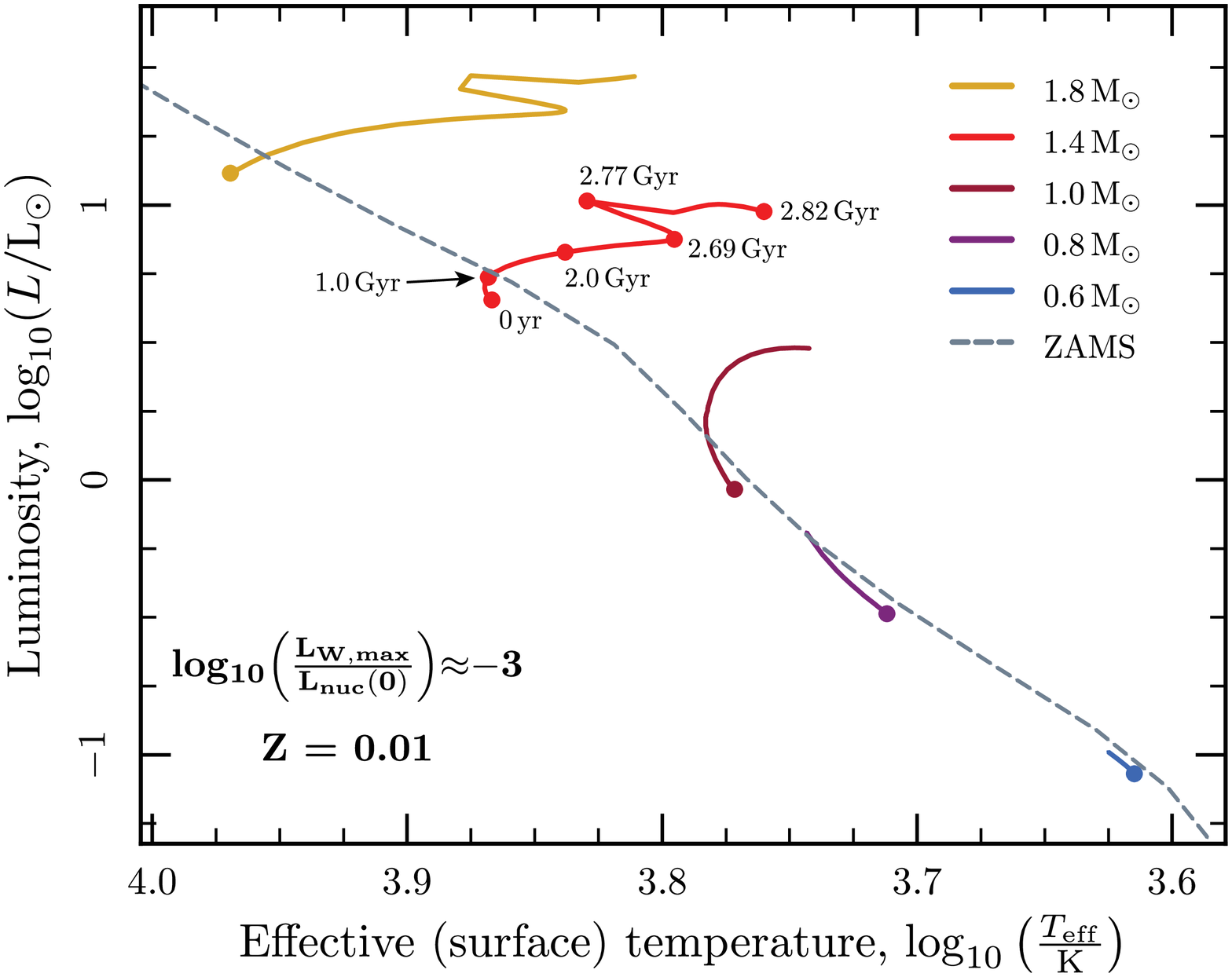} &
\includegraphics[scale=0.25]{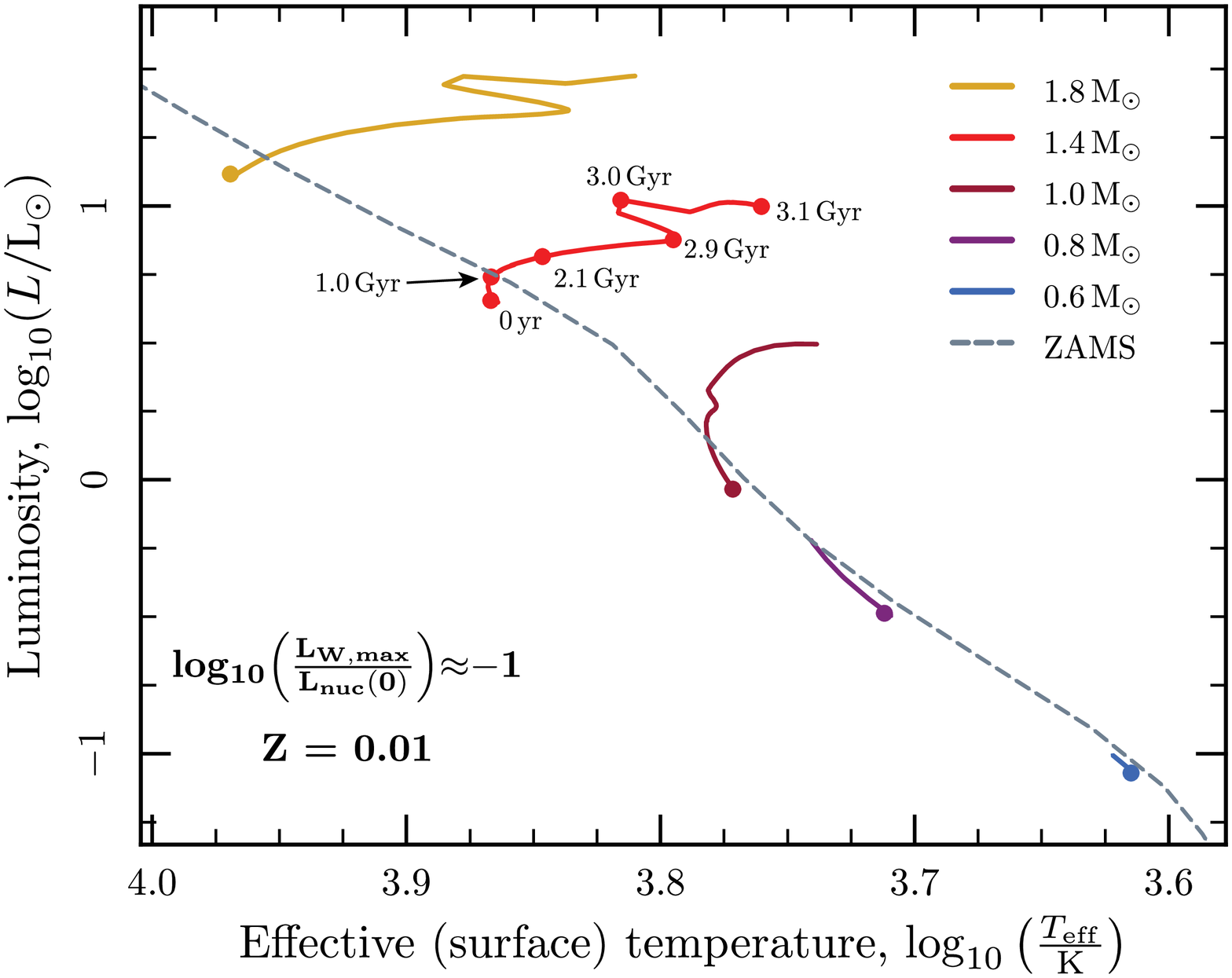} \\
\includegraphics[scale=0.25]{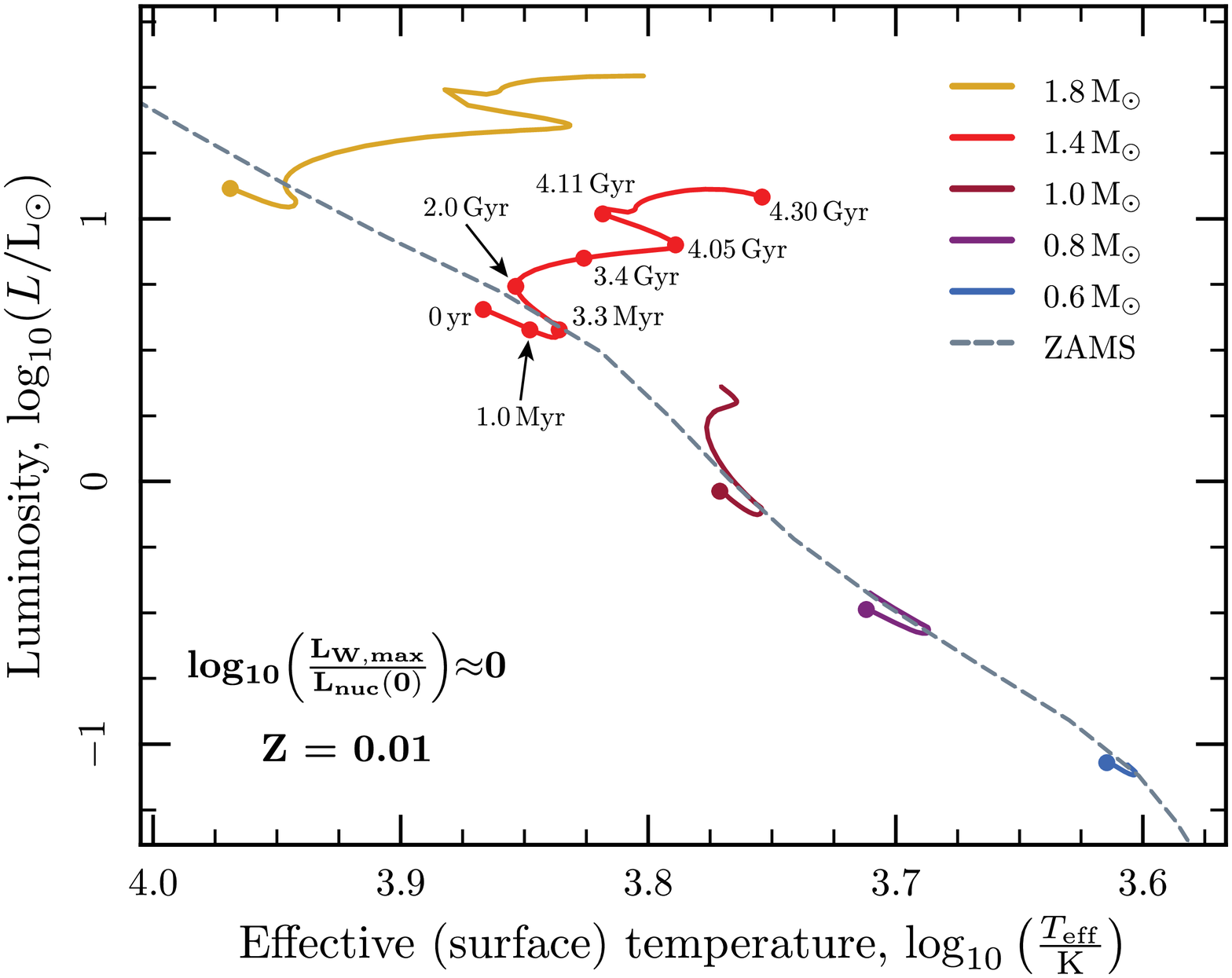} &
\includegraphics[scale=0.25]{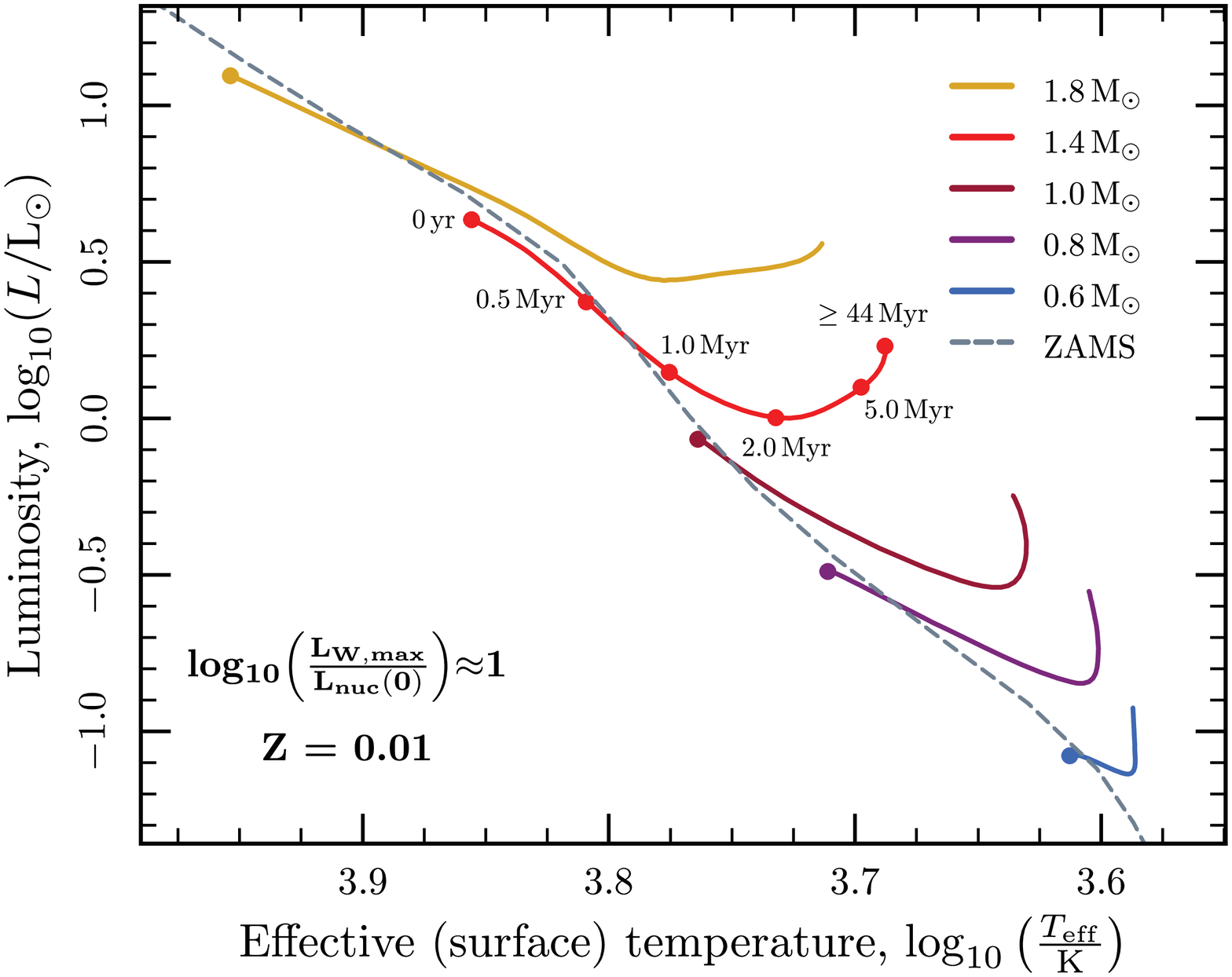} \\
\end{array}$
\end{center}
\caption{HR diagram showing the evolutionary tracks followed by stars of different masses, as WIMPs are allowed to provide different fractions of the stars' total energy budgets.  The energy provided by WIMP annihilation is shown in the bottom left of each sub-plot as the ratio of the maximum luminosity achieved by WIMP annihilation to the initial luminosity due to fusion.  Starting points of tracks are indicated with filled, unlabelled circles, whilst labelled circles give indicative ages during the evolution of 1.4\,M$_\odot$ stars.  Simulations have been halted when the star exhausts its core hydrogen supply or reaches the current age of the Universe.  Stars with a greater luminosity contribution from WIMPs push further up the Hayashi track and spend longer there before returning to the main sequence.  Those which come to be entirely dominated by WIMP annihilation (\emph{bottom right}) evolve back up the Hayashi track on the thermal timescale and halt, holding their position well beyond the age of the 
Universe. (Figure taken from Ref.\cite{SFE09} with kind permission of authors. Reproduced by permission from Oxford University
Press on behalf of the Royal Astronomical Society.)}
\label{Fig_DS_today}
\end{figure}

 High DM densities are the territories in which dark stars may lurk.  DM densities at the center of 
our galaxy provide a refuge ``safe haven" for dark stars.  The dark star menagerie has evolved and diversified.   
Dark stars today are not the supermassive stars of the past, but instead exist on a much reduced scale compared to their earlier 
glory due to the lower DM densities found at the galactic center compared to halos at high redshift.
 Only DSs on the order of a solar mass or less can live on such reduced food supplies.  In the Introduction, we already 
 mentioned the earliest references on dark stars in the current Universe. 
 
 As with DSs in the early Universe,  DSs at the galactic center look a bit different from normal stars unaffected by 
 DM heating.  
 We summarize very briefly the changes made by DM heating on stars at the galactic center; see Ref.
 \cite{SS89, Scott:jea, SFE09, CL09} for more information. Also, Ref.\cite{CL11b} studied the effect of DM on stellar clusters,
 while the use of asteroseismology to detect singatures in the stars produced by the presence of DM has been studied in Ref.
 \cite{CL11a, CL13}.
 We also note that the authors of Ref.\cite{Scott:jea} have written the DarkStars code: a publicly available stellar
 evolution package taking DM heating in stars into account.
 Let us take as the starting point a hydrogen burning star and examine the changes as DM heating also becomes important.
 Stars at the galactic center capture DM via scattering.
 As more DM is captured, DM heating begins to power the star.  The negative
 heat capacity of a star causes it to expand and cool. Fusion becomes less important as
 the central temperature and density of the star drop. The star
 can reverse its  descent onto the main sequence and instead
 traverses up the Hayashi track.
 Figure \ref{Fig_DS_today} 
 illustrates the evolutionary tracks of stars of different
 masses, as WIMPs provide varying fractions of the stars' total energy budgets.
 DSs at the galactic center observationally would look like young stars. With a sufficient amount of DM heating,
 fusion shuts off completely.   With a limitless supply of DM, the star's life can be extended
 indefinitely, at least much longer than if powered by fusion.
 Hence possibly both the first and last stars of the Universe might well be Dark Stars.
 
A second breed of DSs dubbed ``WIMP burners" can live off DM as well, see Ref.\cite{MW07,bertone}.
WIMP burners are degenerate stars (white dwarves or neutron stars).
At the galactic center, there is a large number of white dwarves and neutron stars. Many
of these stars, which would otherwise become fainter with time, can be warmed up by DM heating.  
While the rest of stars continue to evolve, white dwarves and neutron stars  
will appear anomalously hot and young due to the DM heating.

Dark stars at the galactic center are potentially observable.
The galactic center despite being nearby is shrouded in dust. 
In order to observe stars at the galactic center requires going deep into the near  infrared,
which will be possible with the next generation of telescopes.  
The Thirty-Meter Telescope TMT, which will turn on in the next few years, 
can observe stars up to magnitude 22 in the K band which corresponds to stars with sub-solar masses~\cite{0004-637X-622-2-878};
these could be affected by DM heating. JWST could also detect sub-solar mass stars at the galactic center.
If DM affects stars at the galactic center, then the low-mass end of the main sequence would develop a bump
as the low-mass stars are pushed up the Hayashi track. The high-mass end of the main sequence will be unaffected by DM
heating since the energy generated  by DM heating will be small compared to energy generated via fusion.
In the case of degenerate stars, white dwarves will also look anomalously hot compared to the age of stars at the galactic center.
Hence, the galactic center offers a good hunting ground for DSs.

\section{Summary}
\label{sec:summary}

 Dark stars are stars made (almost entirely) of hydrogen and helium but powered by dark matter (DM) annihilation,
 rather than by fusion.  They are in hydrostatic and thermal equilibrium, but with an unusual power source.
 We discussed dark stars (DSs) throughout the history of the Universe, both at early times and today.
 Our focus was the dark stars that may be the first stars to form in the Universe.
 We have reviewed how they come into existence, 
how they grow as long as dark matter fuel persists, and their stellar structure and evolution.   
The studies were done in two different ways, first assuming polytropic interiors and more recently using the MESA stellar evolution code;
the basic results are the same. The structure and evolution of DSs can be seen in Figures 2-5.  DSs are giant, puffy ($\sim$ 10 AU) and
cool (surface temperatures $\sim$10,000 K) objects.  They initially weigh about $\sim 1 M_\odot$, and they grow via accretion from the
surrounding material. As long as they have DM fuel, their surface temperatures remain cool enough that they can keep growing; 
they do not produce
ionizing photons which prevent further accretion.  
Some dark stars may grow to be supermassive dark stars, even reaching masses $> 10^6 \msun$ and luminosities $>10^9 L_\odot$.

It is interesting to speculate that the Initial Mass Function of the first fusion-powered stars may be 
determined by the details of the dark matter distribution resulting from cosmological structure formation.  
 Individual DSs in different halos may end up with a
variety of different masses, depending on the details of the evolution
of the halos they live in. Then, the final DS masses determine the initial masses of the standard Population III stars.
  Once the dark matter fuel inside the DS is exhausted, 
then the star contracts to maintain pressure support.  Lighter DSs become fusion-powered.  On the other end of
the spectrum, the most massive dark stars may collapse directly to supermassive black holes (BHs)
with masses of $> 10^5 M_{\odot}$ without any fusion phase at all, thereby providing seed BHs 
for the many supermassive BHs found throughout the Universe. Thus, the initial masses of
the first fusion-powered stars may depend on the DM environment within which the earlier DSs grow.  

Supermassive dark stars would be detectable in upcoming James Webb Space Telescope observations.  We have shown
their spectra and signatures in JWST. In addition, a new research direction is
the study of pulsations in DSs which lead to variability in their light output. Such variability could be used
to tackle the question of how one can differentiate a DS from an early galaxy. 
Initial investigations found a variety of pulsation periods, including some
which are of the order of months in the observer's frame. If the pulsations are detectable,
DSs may in principle someday be used as novel standard
candles for cosmological studies. It will also be possible to learn about or even possibly 
discover WIMP dark matter by observing the properties of DSs.  

We mention briefly an interesting speculation about another way to detect supermassive DSs.  Black holes emit gravitational waves 
when they coalesce in a binary merger. BHs are point masses relative to DSs. If DSs do grow to be quite massive, their mergers 
should also yield a gravitational wave signature. The fact that they are extended should make their gravitational wave signature
distinguishable from that of supermassive BHs of similar masses. Could this be a potentially interesting way to detect DSs with LIGO and 
maybe with eLISA? Furthermore, they are composed of relatively cool baryons, so the electromagnetic signature of the event 
should also be possibly quite 
distinct from that of a much hotter accretion disk. 

The idea that a new type of star may be discovered in the near future is very exciting.

\ack
We would like to thank our collaborators on the papers reviewed in this paper:
Anthony Aguirre, Peter Bodenheimer, J\"urg Diemand, Paolo Gondolo, Ilian Iliev, Cosmin Ilie, Mike Montgomery, Bill Paxton, Eddie Ruiz, Pearl Sandick, 
Jerry Sellwood, Paul Shapiro, and Don Winget.
 K.F. and T.R.D. acknowledge support from the U.S. Department of Energy under grant DOE-FG02-95ER40899 and the Michigan Center for 
 Theoretical Physics at the University of Michigan. M.V. acknowledges support from the U.S. National Science Foundation 
 via grant AST-0908346.  D.S.\ and K.F.\ are grateful for financial support from the Swedish Research Council (VR) through the Oskar Klein Centre and Stockholm University.

\newpage
\textbf{References} 
\\ 
\\
\bibliography{DS_review}

\end{document}